\documentclass[aps,prb,reprint,groupedaddress,amsmath,amssymb]{revtex4-2}

\usepackage[colorlinks=true,linkcolor=blue,citecolor=blue,urlcolor=blue]{hyperref}

\usepackage{graphicx}

\usepackage{stix}

\newcommand{\lcm}{\mathrm{lcm}}
\newcommand{\dtilde}[1]{\tilde{{\tilde{#1}}}}
\newcommand{\smod}{~\mathrm{mod}\,}

\begin{document}

\title{Frozen Phonons on Dense \textbf{q}-Grids Using Minimal Sets of Non-Diagonal Supercells}

\author{Sebastian Riis Thomsen}
\email[]{seb@chem.au.dk}
\affiliation{Department of Chemistry, Aarhus University, Denmark}

\date{\today}

\begin{abstract}
Phonon dynamics determine a vast number of properties in the solid state, such as the heat capacity, thermal expansion, and thermal conductivity. Phonons are obtained through the dynamical matrix, which, in the frozen phonon approach, is calculated as a finite difference derivative of the analytical Hellmann-Feynman force. This approach unfortunately typically suffers from requiring very large supercells, even for modest $\mathbf{q}$-grids. The non-diagonal supercell method remedies this issue by determining the smallest supercell commensurate with a given $\mathbf{q}$-point, drastically reducing the supercell size, but in turn requiring a new supercell for each $\mathbf{q}$-point on the grid. This introduces redundancy, however, since a single non-diagonal supercell may be commensurate with several $\mathbf{q}$-points. In this work I present a method to determine a minimal set of supercells, such that every $\mathbf{q}$-point on the grid is commensurate with at least one supercell. I present an algorithm to solve the problem and prove that it yields the optimal solution, before summarizing the algorithm using pseudocode. The scaling of the number of supercells with the density of the grid is discussed. Numerical results on a variety of systems show the significant computational advantage of the method, which increases with the density of the grid.
\end{abstract}

\maketitle

\section{Introduction}

The dynamics of phonons control a vast number of properties in the solid state. The energy range of a typical phonon causes it to be populated at room temperature, implying that thermal properties are often governed by phonons. This includes heat capacity \cite{phonons}, thermal expansion \cite{thermal_expansion}, and thermal conductivity \cite{thermal_conductivity}, among others. Phonons are furthermore experimentally observable, for example through neutron diffraction techniques \cite{neutron} or Raman \cite{raman} and infrared \cite{ir} spectroscopy.

Phonons are obtained through the dynamical matrix. In real space, this matrix (also known as the force constant matrix) is given by
\begin{equation}\label{eq:dynmat}
	\mathbf{D}(\mathbf{R}_1, \mathbf{R}_2) = \left[ \frac{1}{\sqrt{m_{\kappa_1} m_{\kappa_2}}} \frac{\partial^2 E}{\partial x_{\kappa_1\alpha_1,\mathbf{R}_1} \partial x_{\kappa_2\alpha_2,\mathbf{R}_2}} \right],
\end{equation}
where $E$ is the electronic energy, $m_\kappa$ is the mass of atom $\kappa$, and $x_{\kappa \alpha,\mathbf{R}}$ is the displacement in the Cartesian direction $\alpha$ away from the equilibrium position of atom $\kappa$ in the unit cell at lattice vector $\mathbf{R}$. The derivative is evaluated at equilibrium. The dynamical matrix has the same translational symmetry as the lattice,
\begin{equation}
	\mathbf{D}(\mathbf{R}_1, \mathbf{R}_2) = \mathbf{D}(\mathbf{0}, \mathbf{R}_2 - \mathbf{R}_1) \equiv \mathbf{D}(\mathbf{R}), \quad \mathbf{R} = \mathbf{R}_2 - \mathbf{R}_1.
\end{equation}
Only one dynamical matrix per lattice vector consequently needs to be computed and stored. The real space dynamical matrices are related to the reciprocal space dynamical matrices by Fourier transform,
\begin{equation}
	\mathbf{D}(\mathbf{q}) = \sum_\mathbf{R} e^{i\mathbf{q}\cdot\mathbf{R}} \mathbf{D}(\mathbf{R}).
\end{equation}
The eigenvalues of $\mathbf{D}(\mathbf{q})$ are squares of the phonon energies, and the eigenvectors are the polarizations of the phonons \cite{elph}. A typical phonon calculation obtains the dynamical matrices on an $n_1\times n_2 \times n_3$ grid of $\mathbf{q}$-points, which can be interpolated to arbitrary $\mathbf{q}$-points, and the interpolation converges with a sufficiently dense grid.

The conceptually simplest way to obtain the dynamical matrices is by using a finite difference approach for the derivative in Eq.\ \eqref{eq:dynmat}, a procedure known as the frozen phonon method. Early frozen phonon calculations obtained the dynamical matrix through a finite difference second derivative of the energy. Modern approaches typically calculate the first derivative of the energy analytically using Hellmann-Feynman forces $F_{\kappa\alpha,\mathbf{R}}$, and then obtain the dynamical matrix from a finite difference first derivative of the force,
\begin{equation}\label{eq:force}
	\mathbf{D}(\mathbf{R}) = \left[ -\frac{1}{\sqrt{m_{\kappa_1} m_{\kappa_2}}}\frac{\partial F_{\kappa_2\alpha_2,\mathbf{R}}}{\partial x_{\kappa_1\alpha_1,\mathbf{0}}} \right].
\end{equation}
With analytical forces, the number of electronic structure calculations needed to obtain this matrix is on the order of the number of atoms $N_\text{atom}$ in the primitive unit cell, in contrast to a finite difference second derivative, which scales as $N_\text{atom}^2$. Despite the use of analytical forces, traditional frozen phonon calculations unfortunately suffer from requiring very large supercells in the electronic structure calculation, even for modest $\mathbf{q}$-grids \cite{nondiagonal}.

Moving beyond frozen phonons, the analytical second derivative of the energy may be obtained directly through Density Functional Perturbation Theory (DFPT) \cite{dfpt}, where only the primitive unit cell is required. This facilitates the calculation of dynamical matrices on dense $\mathbf{q}$-grids. Despite the success of DFPT, its availability is limited by difficulty of implementation, in stark contrast to the simplicity of the frozen phonon approach. Being agnostic to the employed electronic structure calculation, frozen phonons are often used for more advanced electronic treatments such as hybrid functionals \cite{hybrid1, hybrid2} or the GW approach for electron correlation \cite{GW1, GW2}. Modern approaches to anharmonic vibrational effects also tend to stray from linear response approaches such as DFPT \cite{sscha, asdm, tdep}, even though higher-order force constants (third and fourth) have been obtainable using linear response for decades \cite{linresp_anharm1, linresp_anharm2, linresp_anharm3}. The frozen phonon approach is therefore still relevant, motivating the search for ways to increase its efficiency.

\begin{figure}
	\centering
	\includegraphics{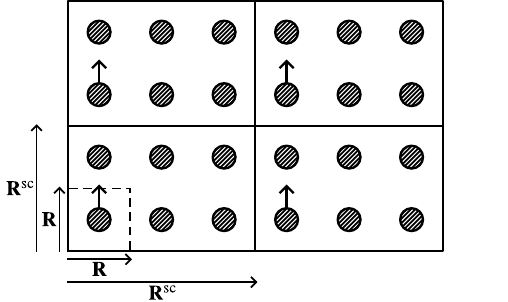}
	\caption{The displacement of an atom in each supercell with Born-von Karman boundary conditions. $\mathbf{R}$ denotes primitive lattice vectors, $\mathbf{R}^\text{sc}$ superlattice vectors.}
	\label{fig:bvk}
\end{figure}

Displacing a single atom in the system breaks the translational symmetry of the lattice, which is detrimental to the performance of solid state electronic structure calculations. This problem is mitigated by calculating the dynamical matrix with Born-von Karman (periodic) boundary conditions $\mathbf{D}_\text{BvK}(\mathbf{R})$ within a supercell, as illustrated in Fig.\ \ref{fig:bvk}. These matrices are periodic with the superlattice vectors $\mathbf{R}^\text{sc}$,
\begin{equation}
	\mathbf{D}_\text{BvK}(\mathbf{R}+\mathbf{R}^\text{sc}) = \mathbf{D}_\text{BvK}(\mathbf{R}).
\end{equation}

It is then possible to calculate $\mathbf{D}(\mathbf{q})$ from $\mathbf{D}_\text{BvK}(\mathbf{R})$ for any $\mathbf{q}$-point which is commensurate with the supercell, such that for any $\mathbf{R}^\text{sc}$,
\begin{equation}
	e^{i\mathbf{q}\cdot\mathbf{R}^\text{sc}} = 1 \quad \Longleftrightarrow \quad \mathbf{q}\cdot\mathbf{R}^\text{sc} = 0 \mod2\pi.
\end{equation}
The modulo operation $a=b\smod c$ signifies that the difference between $a$ and $b$ is a multiple of $c$.

When calculating phonons on an $n_1\times n_2 \times n_3$ grid of $\mathbf{q}$-points, there are different approaches for choosing the supercell. The most straightforward is with a single, so-called \textit{diagonal} supercell, constructed so every $\mathbf{q}$-point is commensurate with this one supercell. The number of atoms in the supercell is
\begin{equation}
	N_\text{atom}^\text{sc} = N_\text{atom}n_1n_2n_3.
\end{equation}
While often used for its simplicity, the number of atoms scales rapidly with the size of the grid, and is usually only feasible for coarse $\mathbf{q}$-grids.

A different approach was introduced by Lloyd-Williams and Monserrat using \textit{non-diagonal} supercells \cite{nondiagonal}. For every $\mathbf{q}$-point on the grid, the smallest supercell commensurate with that $\mathbf{q}$-point is generated. This significantly lowers the number of atoms in each supercell to at most
\begin{equation}
	N_\text{atom}^\text{sc} = N_\text{atom}\,\lcm(n_1, n_2, n_3).
\end{equation}
The function ``$\lcm$'' is the least common multiple of the three integers. While requiring substantially fewer atoms, the number of supercells needed is increased from 1 to $n_1n_2n_3$, as a new supercell is generated for every $\mathbf{q}$-point, though this number may be approximately halved by time-reversal symmetry. The use of non-diagonal supercells presents a computational saving, as shown by Lloyd-Williams and Monserrat, since electronic structure calculations often scale poorly with the number of atoms, making the trade-off of increasing the number of supercells worthwhile.

Using a new supercell for each $\mathbf{q}$-point is superfluous, however, since several $\mathbf{q}$-points may be commensurate with the same supercell. If $\mathbf{q}'$ is related to $\mathbf{q}$ by
\begin{equation}\label{eq:qprime}
	\mathbf{q}' = t\mathbf{q} \mod\mathbf{Q},
\end{equation}
where $t\in\mathbb{Z}$ (i.e.\ $t$ is an integer) and $\mathbf{Q}$ is a reciprocal lattice vector, and if $\mathbf{q}$ is commensurate with a given supercell, then $\mathbf{q}'$ is also commensurate with that supercell,
\begin{equation}
	\mathbf{q}'\cdot\mathbf{R}^\text{sc} = t\left(\mathbf{q}\cdot\mathbf{R}^\text{sc}\right) = 0 \mod2\pi.
\end{equation}
The non-diagonal supercell commensurate with $\mathbf{q}$ can consequently be used to obtain all $\mathbf{D}(\mathbf{q}')$ with $\mathbf{q}'$ of the form Eq.\ \eqref{eq:qprime}. This implies the existence of a minimal set of non-diagonal supercells, such that every $\mathbf{q}$-point on the grid is commensurate with at least one supercell.

This work presents an algorithm to determine the minimal set of non-diagonal supercells needed to obtain the dynamical matrix $\mathbf{D}(\mathbf{q})$ for every $\mathbf{q}$-point on an $n_1 \times n_2 \times n_3$ uniform $\mathbf{q}$-grid. This is referred to as the \textit{cover} method, reflecting that every $\mathbf{q}$-point on the grid is ``covered'' by the minimal set of supercells. It is first proven that the algorithm yields the minimal number of supercells.  The algorithm is then summarized using pseudocode, and a comment is made about the scaling of the number of supercells with the size of the $\mathbf{q}$-grid. Numerical results are presented for aluminium (Al), fluorite (CaF$_2$), graphene (C), cooperite (PtS), and niobium-tin (Nb$_3$Sn), showing the gain in computational efficiency over simply calculating every $\mathbf{q}$-point independently.

\section{Theory}

On an $n_1 \times n_2 \times n_3$ $\mathbf{q}$-grid, each $\mathbf{q}$-point can be written in fractional coordinates as
\begin{equation}
	\mathbf{q} = \begin{pmatrix}
		m_1 / n_1 \\
		m_2 / n_2 \\
		m_3 / n_3
	\end{pmatrix}, \quad
	m_i = 0,1,\dots,n_i-1.
\end{equation}
Keeping the size of the grid implicit, it is more convenient to express the $\mathbf{q}$-point as $\mathbf{q} = (m_1, m_2, m_3)$. Two points $\mathbf{q}=(m_1, m_2, m_3)$ and $\mathbf{q}'=(m_1', m_2', m_3')$ are equal modulo a reciprocal lattice vector if $m_i = m_i' \smod n_i$ for $i=1,2,3$. The integer multiples of $\mathbf{q}$ are
\begin{equation}
		\mathbf{q}' = t\mathbf{q} = (tm_1, tm_2, tm_3) \mod (n_1, n_2, n_3)
\end{equation}
for $t \in \mathbb{Z}$. Consider the set of all $\mathbf{q}'$ related to $\mathbf{q}$ this way. The set certainly includes $\mathbf{0}=(0, 0, 0)$, for example by $t=0$. The set also includes a $\mathbf{q}''$ for every $\mathbf{q}'$ such that $\mathbf{q}'+\mathbf{q}''=\mathbf{0}\smod (n_1,n_2,n_3)$, i.e.\ the additive inverse of $\mathbf{q}'$. This set is thereby a \textit{group} under addition, and is said to be \textit{generated} by $\mathbf{q}$, written as
\begin{equation}
	\langle \mathbf{q} \rangle_{(n_1, n_2, n_3)} \equiv \{ t\mathbf{q}~\smod (n_1,n_2,n_3)~:~t\in\mathbb{Z}\}.
\end{equation}
The entire grid also forms a group,
\begin{equation}
	\mathbb{Z}_{n_1} \times \mathbb{Z}_{n_2} \times \mathbb{Z}_{n_3} = \{ (m_1, m_2, m_3)~:~m_i\in\mathbb{Z}_{n_i} \}
\end{equation}
with $\mathbb{Z}_n=\{0,1,\dots,n-1\}$ denoting the group of integers modulo $n$ under addition. Any $\mathbf{q}'\in \langle \mathbf{q} \rangle_{(n_1,n_2,n_3)}$ will be commensurate with any supercell $\mathbf{q}$ is commensurate with, as discussed above. The task at hand is then to determine a minimal set $\mathcal{Q}$ of $\mathbf{q}$-points, such that the union of all $\langle \mathbf{q} \rangle_{(n_1,n_2,n_3)}$ covers the entire grid,
\begin{equation}\label{eq:cyclic_cover}
	\mathbb{Z}_{n_1} \times \mathbb{Z}_{n_2} \times \mathbb{Z}_{n_3} = \bigcup_{\mathbf{q}\in\mathcal{Q}} \langle \mathbf{q} \rangle_{(n_1, n_2, n_3)}.
\end{equation}
This is often called a \textit{set cover} \cite{setcover}, hence the name of the cover method. Once the set cover is determined, the minimal required set of non-diagonal supercells contains one supercell for each $\mathbf{q}$ in $\mathcal{Q}$.

The algorithm described below gradually adds $\mathbf{q}$-points to $\mathcal{Q}$ until Eq.\ \eqref{eq:cyclic_cover} holds. It will often be convenient to consider whether a given point $\mathbf{q}'$ is included in the union built up so far in the algorithm. The point is said to be \textit{covered} whenever that is the case, and specifically $\mathbf{q}'$ is \textit{covered by} $\mathbf{q}$ if $\mathbf{q}'\in\langle \mathbf{q} \rangle_{(n_1, n_2, n_3)}$.

The gradual buildup of $\mathcal{Q}$ means that the algorithm sometimes determines a $\mathbf{q}'$ to add which is already covered by some other $\mathbf{q}\in\mathcal{Q}$. The addition of $\mathbf{q'}$ is then unnecessary, as any $\mathbf{q}''$ covered by $\mathbf{q}'$ is also covered by $\mathbf{q}$. When building up the union in Eq.\ \eqref{eq:cyclic_cover} along with $\mathcal{Q}$, this may be avoided simply by checking whether $\mathbf{q'}$ is already in the union.

The logic of the algorithm is best appreciated by starting with a one-dimensional grid, advancing to two dimensions, and then finalizing in three dimensions. 

\subsection{One Dimension}\label{sect:1D}

\begin{figure}
	\centering
	\includegraphics{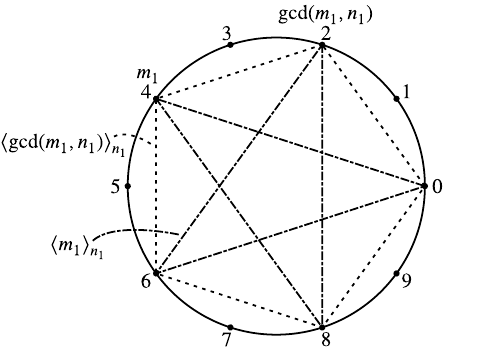}
	\caption{Visualization of a grid as points on a circle. In this example, $n_1=10$, $m_1=4$, so $\gcd(m_1, n_1)=2$. The groups $\langle m_1 \rangle_{n_1}$ and $\langle \gcd(m_1, n_1) \rangle_{n_1}$ are shown, with consecutive group elements connected by lines. The groups contain the same elements, but in different orders.}
	\label{fig:1d_circle}
\end{figure}

The $\mathbf{q}$-points on a one-dimensional grid with size $n_1$ are given by a single integer, $\mathbf{q}=m_1\in\mathbb{Z}_{n_1}$. The grid is periodic with period $n_1$, thereby lending itself to visualization as a circle, with the $\mathbf{q}$-points distributed around it, as shown in Fig.\ \ref{fig:1d_circle}. Consider a particular $\mathbf{q}=m_1$ and the generated group $\langle m_1 \rangle_{n_1}$. Figure \ref{fig:1d_circle} shows the elements $m_1'=tm_1~\mathrm{mod}\,n_1$ of the group, connected by lines in the order they appear, as $t$ is increased from $1$ to $n_1$. The smallest non-zero integer in $\langle m_1 \rangle_{n_1}$ is the greatest common divisor between $m_1$ and $n_1$, denoted $\gcd(m_1, n_1)$, as is shown in the supplemental material \cite{supmat} (SM) Sec.\ SA\,10. Let the notation $a\mid b$ express that $a$ divides $b$. It is always possible to write $m_1$ as a multiple of $\gcd(m_1, n_1)$, since $\gcd(m_1, n_1) \mid m_1$. Any element $m_1'\in\langle m_1 \rangle_{n_1}$ can then be written as
\begin{equation}
	\begin{split}
		m_1' = tm_1 &= t \frac{m_1}{\gcd(m_1, n_1)} \gcd(m_1, n_1) \mod n_1 \\
			&\in \langle \gcd(m_1, n_1) \rangle_{n_1}.
	\end{split}
\end{equation}
The converse is also true: any $m_1'\in \langle \gcd(m_1, n_1) \rangle_{n_1}$ can be written as a multiple of $m_1$. This implies that
\begin{equation}
	\langle m_1 \rangle_{n_1} = \langle \gcd(m_1, n_1) \rangle_{n_1}.
\end{equation}
This relation will be used extensively in the following sections, as the greatest common divisor has several useful properties. The properties used in this work are listed (and proven) in the SM, Sec.\ SA. Though the elements in $\langle m_1 \rangle_{n_1}$ and $\langle \gcd(m_1, n_1) \rangle_{n_1}$ are the same, the ordering may be different. Both groups are illustrated in Figure \ref{fig:1d_circle}, with their respective orderings.

The smallest integer $t$ such that $t\gcd(m_1,n_1)=n_1$ is $t=n_1/\gcd(m_1,n_1)$, as $\gcd(m_1,n_1)$ is a divisor of $n_1$. Increasing $t$ from $1$ to $n_1/\gcd(m_1, n_1)$, $n_1/\gcd(m_1, n_1)$ unique elements of the group $\langle \gcd(m_1, n_1) \rangle_{n_1}$ are encountered, implying that the size of the group is
\begin{equation}
	|\langle m_1 \rangle_{n_1}| = |\langle \gcd(m_1, n_1) \rangle_{n_1}| = \frac{n_1}{\gcd(m_1, n_1)}.
\end{equation}

With the value of $m_1$ chosen in Fig.\ \ref{fig:1d_circle}, clearly not every point in $Z_{n_1}$ is covered. Choosing instead an $m_1$ such that $\gcd(m_1, n_1)=1$, the generated group would be
\begin{equation}
	\langle m_1 \rangle_{n_1} = \langle \gcd(m_1, n_1) \rangle_{n_1} = \langle 1 \rangle_{n_1} = \mathbb{Z}_{n_1}.
\end{equation}
The one-dimensional case thus solves Eq.\ \eqref{eq:cyclic_cover} with a single $\mathbf{q}$-point, chosen such that $\gcd(\mathbf{q}, n_1) = \gcd(m_1, n_1) = 1$. Whenever $\gcd(m_1, n_1)=1$, $m_1$ is said to be \textit{coprime} to $n_1$ and vice versa.

\subsection{Two Dimensions}

\begin{figure}
	\centering
	\includegraphics{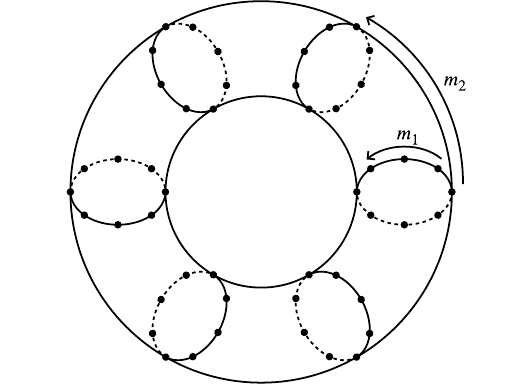}
	\caption{Visualization of $\mathbb{Z}_{n_1}\times\mathbb{Z}_{n_2}$ as a torus. Index $m_2$ moves between the circles of the torus, while $m_1$ moves around the points on each circle. In this example, $n_1 = 8$ and $n_2 = 6$.}
	\label{fig:2d_torus}
\end{figure}

The two-dimensional grid $\mathbb{Z}_{n_1} \times \mathbb{Z}_{n_2}$ is periodic in both dimensions,
\begin{equation}
	(m_1+n_1, m_2) = (m_1, m_2 + n_2) = (m_1, m_2) \smod (n_1, n_2).
\end{equation}
This makes the grid equivalent to a torus, as visualized in Fig.\ \ref{fig:2d_torus}.

\textit{Circle} $m_2$ will be defined as the set of points $\{(m_1, m_2):m_1\in\mathbb{Z}_{n_1}\}$. A circle is said to be \textit{covered} if every point on the circle is covered, and circle $m_2'$ is covered by circle $m_2$ if every point on circle $m_2'$ is covered by a point on circle $m_2$. Circle $m_2$ is similarly said to \textit{reach} $m_2'$ if at least one point on $m_2'$ is covered by a point on $m_2$, i.e.\ $m_2'\in\langle m_2 \rangle_{n_2}$.

The two-dimensional algorithm proceeds in two main steps:
\begin{enumerate}
	\item Assuming circle $m_2$ is covered, deduce which circles $m_2'$ are covered by $m_2$. It is then unnecessary to add any such $m_2'$ to $\mathcal{Q}$. The remaining circles $m_2$, which are not covered by any other circle, contain at least one point each which must be added to $\mathcal{Q}$.
	\item Find the minimal set of points $(m_1, m_2)$ on circle $m_2$ needed to cover the remaining points on $m_2$ (and thereby all the circles $m_2'$ covered by $m_2$).
\end{enumerate}

\subsubsection{Step 1}\label{sect:2d_step_1}

\begin{figure}
	\centering
	\includegraphics{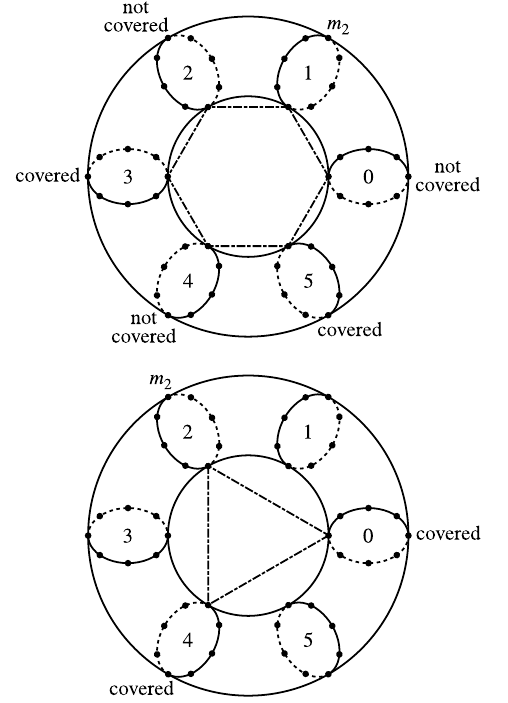}
	\caption{For a given divisor $m_2$ of $\pi_{n_1}(n_2)$, circle $m_2'$ is covered if Eq.\ \eqref{eq:cover_condition} is satisfied. The two tori correspond to two different divisors. The dash-dotted line in the middle indicates which circles are reached by $m_2$, while the surrounding text indicates whether they are covered by circle $m_2$. In this example $n_1=8$ and $n_2=6$ so $\pi_{n_1}(n_2) = 2$, i.e.\ $6=2\cdot3$ and $\gcd(3, 8)=1$. The divisors of 2 are $\{1, 2\}$. Every circle is covered by either $m_2=1$ or $m_2=2$, but none are covered by both.}
	\label{fig:2d_step_1}
\end{figure}

Assume $m_2'$ can be reached by $m_2$. Starting at a point $(m_1, m_2)$, there exists some integer $k$ such that $(m_1',m_2') = (km_1, km_2)\smod(n_1, n_2)$. As $k$ is increased from 1 to $|\langle m_2 \rangle_{n_2}| =n_2 / \gcd(m_2, n_2)$, every circle $m_2'$ reached by circle $m_2$ is encountered once. This forms a closed loop across the torus, as illustrated in Fig.\ \ref{fig:2d_step_1} for two different values of $m_2$. Even though $m_2$ reaches $m_2'$, it does not necessarily cover $m_2'$, as is illustrated in the figure as well.

For two integers $a$ and $b$, $a$ can be written uniquely as $a = vr$, where $r$ is the largest divisor of $a$ coprime to $b$. The function $\pi_b(a)$ is then defined as $\pi_b(a)\equiv v$. Any integer coprime to $b$ is also coprime to $\pi_b(a)$ (see Sec.\ SA\,13 of the SM) and any divisor of $\pi_b(a)$ also divides $a$.

For every divisor $d$ of $\pi_{n_1}(n_2)$, at least one $\mathbf{q}$-point with $m_2=d$ needs to be included in $\mathcal{Q}$, and no $\mathbf{q}$-points where $m_2$ does not divide $\pi_{n_1}(n_2)$. The proof of this is given in Sec.\ SB\,1 of the SM, where it is shown that $m_2$ covers $m_2'=km_2$ if and only if
\begin{equation}\label{eq:cover_condition}
	\gcd\left(k,\frac{n_2}{\gcd(m_2, n_2)}, n_1\right) = 1.
\end{equation}
This yields exactly those values of $m_2$ which need to be present in $\mathcal{Q}$.

 \subsubsection{Step 2}\label{sect:2d_step_2}

\begin{figure}
	\centering
	\includegraphics{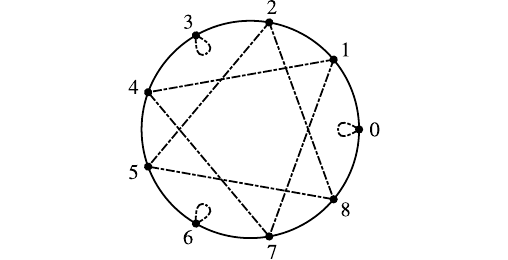}
	\caption{An illustration of the different sets $\mathcal{S}$, chosen to be as large as possible. Each polygon inside the circle represents a unique $\mathcal{S}$ with the ``raindrop'' shapes representing sets with only one element. In this example $n_1 = 9$, $\gcd(\tilde n_2, n_1) = 3$, and $v=3$, so the possible values of $d$ in Eq.\ \eqref{eq:d_relates_to_mprime} are 1 and 3. There are two sets with $d=1$ and three with $d=3$. Every point is included in one and only one $\mathcal{S}$.}
	\label{fig:2d_step_2}
\end{figure}

 Consider a specific circle $m_2$. It was established above that only those $m_2$ which divide $\pi_{n_1}(n_2)$ need to be included in $\mathcal{Q}$, so this is assumed to be the case. Then $m_2$ is a divisor of $n_2$, so $\gcd(m_2, n_2)=m_2$. Starting at $(m_1, m_2)$ and looping around the torus $t$ times as described above, the resulting point $(m_1', m_2)$ has the first coordinate
\begin{equation}
		m_1' = m_1\left(1+t\frac{n_2}{m_2}\right) = m_1\left(1+t\tilde n_2\right) \mod n_1
\end{equation}
with $\tilde n_2 \equiv \frac{n_2}{m_2}$. As the torus is looped around several times, the starting point covers all $(m_1', m_2)$, with $m_1'$ taking the values 
\begin{equation}\label{eq:S_def}
		m_1' \in m_1 + \langle m_1 \tilde n_2 \rangle_{n_1} = m_1 + \langle \gcd(m_1 \tilde n_2, n_1) \rangle_{n_1} \smod n_1.
\end{equation}
Denote this set $\mathcal{S}$. Figure \ref{fig:2d_step_2} illustrates such sets for a few different starting points $(m_1, m_2)$. The point $(m_1, m_2)$ covers all the other $(m_1', m_2)$ for $m_1'\in\mathcal{S}$. Choosing instead $(m_1',m_2)$ as the starting point and following the procedure outlined above, the resulting set may only be a subset of $\mathcal{S}$,
\begin{equation}
	m_1' + \langle m_1'\tilde n_2 \rangle_{n_1} \subseteq \mathcal{S}.
\end{equation}
It is therefore not sufficient to start from an arbitrary $m_1'\in\mathcal{S}$ since $(m_1', m_2)$ does not necessarily cover the remaining points $(m_1'', m_2)$ with $m_1'' \in \mathcal{S}$. 

From the definition of $\mathcal{S}$ in Eq.\ \eqref{eq:S_def}, the smallest non-zero difference between two elements in $\mathcal{S}$ must be $\gcd(m_1 \tilde n_2, n_1) = d\gcd(\tilde n_2, n_1)$ for some integer $d$. Every $m_1'$ on circle $m_2$ is included in at least one such set $\mathcal{S}$. When it is included in more than one, the largest $\mathcal{S}$ must be identified, to minimize the number of sets needed to cover the entire circle. The largest such set $\mathcal{S}$ which includes $m_1'$ has
\begin{equation}\label{eq:d_relates_to_mprime}
	\begin{gathered}
		d = \gcd\left( m_1', \pi_{\tilde n_2}\left(\frac{n_1}{\gcd(\tilde n_2, n_1)}\right) \right) = \gcd\left( m_1', v \right), \\
		v \equiv \pi_{\tilde n_2}\left(\frac{n_1}{\gcd(\tilde n_2, n_1)}\right).
	\end{gathered}
\end{equation}
For every divisor $d$ of $v$, there exists at least one such $\mathcal{S}$. The proof of this is given in Sec.\ SB\,2 of the SM. The sets illustrated in Fig.\ \ref{fig:2d_step_2} are chosen to be as large as possible.

To find the sets $\mathcal{S}$, note that every $m_1'\in\mathcal{S}$ is divisible by $d$, and that there is exactly one $m_1'\in\mathcal{S}$ between $d$ and $d\gcd(\tilde n_2, n_1)$, since the smallest non-zero difference between two elements in $\mathcal{S}$ is $d\gcd(\tilde n_2, n_1)$. Then search for an $m_1'$ in the range $d$ to $d\gcd(\tilde n_2, n_1)$ in steps of $d$, until $m_1'$ satisfies Eq.\ \eqref{eq:d_relates_to_mprime}. Each such $m_1'$ is part of a unique set $\mathcal{S}$, within which must then be found an $m_1$ satisfying
\begin{equation}
	\gcd\left( m_1 \tilde n_2, n_1 \right) = d\gcd\left(\tilde n_2, n_1 \right).
\end{equation}
The resulting $\mathbf{q} = (m_1, m_2)$ covers all other points $(m_1',m_2)$ with $m_1'\in\mathcal{S}$, and is then added to $\mathcal{Q}$, if it is not already included in the union of $\langle \mathbf{q}' \rangle_{(n_1,n_2)}$ for $\mathbf{q}'\in\mathcal{Q}$ (Eq.\ \eqref{eq:cyclic_cover}) built up so far. This provides the recipe for constructing $\mathcal{Q}$, thereby concluding the solution for the two-dimensional grid.

\subsection{Three Dimensions}

\begin{figure}
	\centering
	\includegraphics{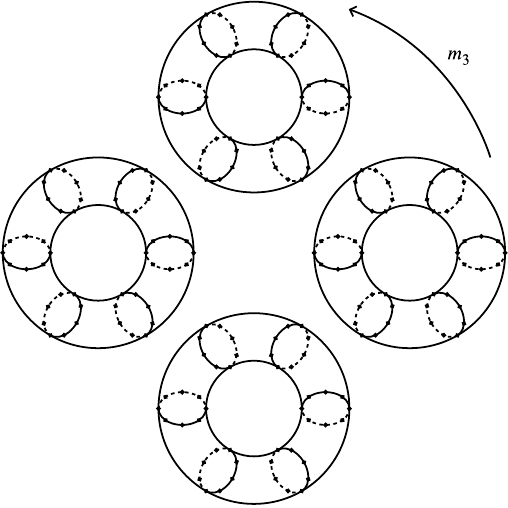}
	\caption{The three-dimensional grid is equivalent to a circle of tori, where the third index specifies a torus in that circle. In this example $n_1=8$, $n_2=6$, and $n_3=4$.}
	\label{fig:3d_tori}
\end{figure}

The three-dimensional grid $\mathbb{Z}_{n_1} \times \mathbb{Z}_{n_2} \times \mathbb{Z}_{n_3}$ is periodic in three dimensions, which can be visualized as a circle of tori, as seen in Fig.\ \ref{fig:3d_tori}. The third coordinate $m_3$ specifies a given torus. Extending the language from two dimensions, the torus $m_3$ is the set of points $\{(m_1, m_2, m_3):m_1\in\mathbb{Z}_{n_1},m_2\in\mathbb{Z}_{n_2}\}$. Torus $m_3'$ is \textit{covered} by torus $m_3$ if every circle on $m_3'$ is covered by the circles on $m_3$. Torus $m_3$ reaches torus $m_3'$ if $m_3'\in\langle m_3 \rangle_{n_3}$.

The three-dimensional algorithm consists of four steps:
\begin{enumerate}
	\item Assuming torus $m_3$ is covered, find the other tori $m_3'$ which are covered by $m_3$. This yields the values of $m_3$ for which a $\mathbf{q}$-point with $m_3$ as the third coordinate needs to be included in $\mathcal{Q}$.
	\item On a given torus $m_3$, identify which circles $m_2'$ on this torus are reached by which other circles $m_2$. This divides the torus into sets $\mathcal{S}$ of circles, analogous to the division into sets of points in Fig.\ \ref{fig:2d_step_2}.
	\item Considering a given set of circles $\mathcal{S}$, a circle $m_2'\in\mathcal{S}$ may be reached by another circle $m_2\in\mathcal{S}$, without being covered by $m_2$. An additional subdivision of $\mathcal{S}$ reveals the circles $m_2$ which need to be covered, such that the remaining circles $m_2'$ are covered by one of the $m_2$ circles. This yields the values of $m_2$ which need to be present in at least one of the $\mathbf{q}\in\mathcal{Q}$.
	\item Given a circle $m_2$ on torus $m_3$, the points on the circle are divided into sets, again as in Fig.\ \ref{fig:2d_step_2}. In each such set, a given $m_1$ which covers all the other $m_1'$ in the set is identified, and the final $\mathbf{q}=(m_1,m_2,m_3)$ is added to $\mathcal{Q}$.
\end{enumerate}

\subsubsection{Step 1}

\begin{figure}
	\centering
	\includegraphics{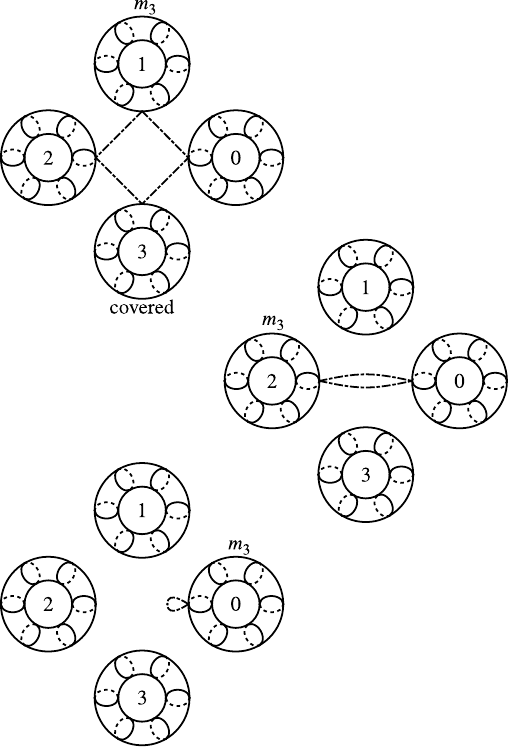}
	\caption{A circle of tori for each divisor of $v=\pi_{n_1n_2}(n_3)$. The polygon in the middle of the tori indicates which tori are reached by $m_3$, and the text indicates if they are covered by $m_3$, with a lack of text indicating that they are not. In this example $n_2=6$ and $n_3=4$, so $v=\pi_{n_1n_2}(n_3)=4$, regardless of $n_1$. The divisors of $v$ are 1 (corresponding to the top circle of tori), 2 (the middle circle of tori), and 4 (the bottom circle of tori).}
	\label{fig:3d_step_1}
\end{figure}

A given torus $m_3$ reaches the tori $m_3'$ where $m_3' \in \langle m_3 \rangle_{n_3}$, though it does not necessarily cover them.
Let $v=\pi_{n_1n_2}(n_3)$. Torus $m_3'$ is then covered by torus $m_3 = \gcd(m_3', v)$ and not by any other divisor of $v$, and any $m_3$ which does not divide $v$ covers the same circles $m_3'$ as some divisor of $v$. This implies that for every divisor $m_3$ of $v$, at least one $\mathbf{q}\in\mathcal{Q}$ must have $m_3$ as its third coordinate, and that it is unnecessary to have any $\mathbf{q}\in\mathcal{Q}$ where the third coordinate is not a divisor of $v$. Figure \ref{fig:3d_step_1} illustrates the circle of tori, showing which tori are reached and which are covered by the different divisors $m_3$ of $v$. The proof is given in Sec.\ SC\,1 of the SM.

\subsubsection{Step 2}

Consider the starting point $(m_1, m_2, m_3)$, where $m_3$ is a divisor of $\pi_{n_1 n_2}(n_3)$. Let $\tilde n_3 = n_3 / m_3$. Looping around the circle of tori takes $\tilde n_3$ multiples of $(m_1, m_2, m_3)$. Looping $t$ times, the starting point becomes $(m_1', m_2', m_3)$, with
\begin{equation}
	m_i' = m_i (1+t \tilde n_3) = m_i + t m_i \tilde n_3 \mod n_i,
\end{equation}
for $i=1,2$. The circles on torus $m_3$ are divided into sets $\mathcal{S}$ of the form
\begin{equation}
	\mathcal{S} = m_2 + \langle m_2 \tilde n_3 \rangle_{n_2} \mod n_2.
\end{equation}
The smallest non-zero difference between two elements in a given set is $\gcd(m_2\tilde n_3, n_2) = d\gcd(\tilde n_3, n_2)$ for some integer $d$, and a circle $m_2'$ is part of such a set if
\begin{equation}\label{eq:m2p_condition}
	\gcd\left(m_2', d\gcd(\tilde n_3, n_2), \frac{n_2}{\gcd(\tilde n_3, n_2)}\right) = d.
\end{equation}

\begin{figure}
	\centering
	\includegraphics{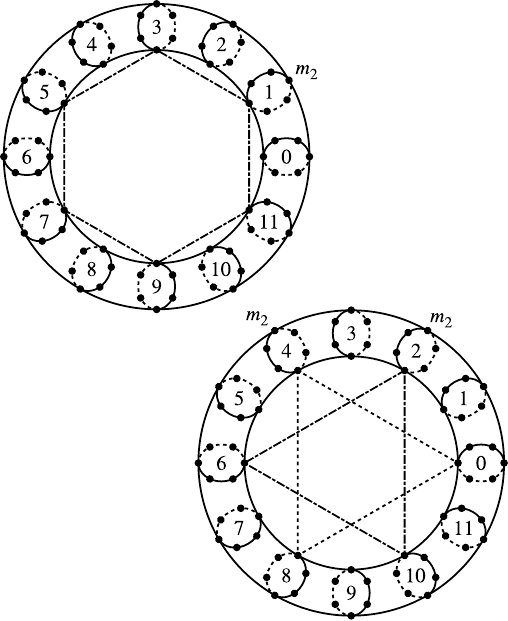}
	\caption{The circles on the torus are divided into sets $\mathcal{S}$, as shown with polygons connecting the elements in the sets. The smallest non-zero difference between two elements of $\mathcal{S}$ is $d \gcd(\tilde n_3, n_2)$, where $d$ is a divisor of $\pi_{\tilde n_3}(n_2 / \gcd(\tilde n_3, n_2))$. In this example $n_1 = 6$, $n_2 = 12$, $\tilde n_3 = 10$. Then $\gcd(\tilde n_3, n_2)=2$ and $\pi_{\tilde n_3}(n_2 / \gcd(\tilde n_3, n_2)) = 2$, the divisors of which are 1 and 2. The top torus shows the single set $\mathcal{S}$ with $d=1$, and the bottom torus shows the two sets with $d=2$. A circle $m_2$ is indicated for each set which reaches the other circles in the set.}
	\label{fig:3d_step_2}
\end{figure}

Let $v=\pi_{\tilde n_3}( n_2/\gcd(\tilde n_3, n_2))$. For every divisor $d$ of $v$, at least one set $\mathcal{S}$ exists where the smallest non-zero difference between elements is $d\gcd(\tilde n_3, n_2)$. An element $m_2$ in each such set can then be identified by the requirement $\gcd(m_2, n_2/\gcd(\tilde n_3, n_2)) = d$, which reaches all other $m_2'\in\mathcal{S}$. Figure \ref{fig:3d_step_2} shows a torus, with the circles on the torus divided into sets. An $m_2$ is shown for each set which reaches all other $m_2'$ in the set. The proof is given in Sec.\ SC\,2 of the SM.

The sets $\mathcal{S}$ may be found as in Sec.\ \ref{sect:2d_step_2} by searching for an $m_2'$ from $d$ to $d\gcd(\tilde n_3, n_2)$ in steps of $d$. Every $m_2'$ which satisfies 
\begin{equation}
	\gcd(m_2', v) = d
\end{equation}
is part of a unique $\mathcal{S}$. This is done for every divisor $d$ of $v$.

\subsubsection{Step 3}

\begin{figure}
	\centering
	\includegraphics{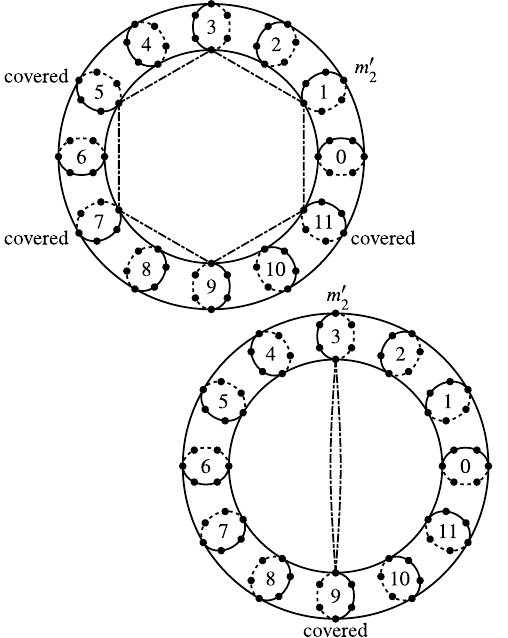}
	\caption{The set of circles $\mathcal{S}$ contains a subset $\mathcal{T}\subseteq\mathcal{S}$ for every divisor $e$ of $w$ (Eq.\ \eqref{eq:w_definition}). Some $m_2''\in\mathcal{T}$ are covered by a circle $m_2'\in\mathcal{T}$, while others are not -- these others are covered in another subset. In this example, $n_1 = 6$, $n_2 = 12$, and $n_3 = 10$. The $\mathcal{S}$ shown here has $d=1$, so $w=3$, the divisors of which are 1 (the top torus, where $\mathcal{T}=\mathcal{S}$) and 3 (the bottom torus). Every circle in $\mathcal{S}$ is covered in exactly one $\mathcal{T}$.}
	\label{fig:3d_step_3}
\end{figure}

The sets $\mathcal{S}$ identified in the previous step indicate all the $m_2'$ which may be reached by the same $m_2$. This does not, however, imply that all the circles $m_2'\in\mathcal{S}$ are covered by circle $m_2$. Let $d$ be the divisor of $v$ corresponding to the given set $\mathcal{S}$, and let $\tilde n_2 = n_2 / d$. Then define
\begin{equation}\label{eq:w_definition}
	w = \pi_{n_1} \left( \frac{\tilde n_2}{\pi_{\tilde n_3}(\tilde n_2)} \right).
\end{equation}
For every divisor $e$ of $w$, a unique subset $\mathcal{T}\subseteq\mathcal{S}$ exists, defined by 
\begin{equation}
	\mathcal{T} = \left\{ m_2'\in\mathcal{S} : ed \mid m_2' \right\}.
\end{equation}
The smallest non-zero difference between two elements of $\mathcal{T}$ is $ed\gcd(\tilde n_3, n_2)$. The circles $m_2''\in\mathcal{T}$ for which
\begin{equation}
	\gcd\left( m_2'', \frac{n_2}{\gcd(\tilde n_3, n_2)} \right) = edr
\end{equation}
with $\gcd(r, n_1)=1$, are covered by an $m_2'\in\mathcal{T}$ with
\begin{equation}
	\gcd\left( m_2', \frac{n_2}{\gcd(\tilde n_3, n_2)} \right) = ed.
\end{equation}
This $m_2'$ must be identified. Every $m_2''\in\mathcal{S}$ is covered in exactly one $\mathcal{T}$ by $m_2'\in\mathcal{T}$. Figure \ref{fig:3d_step_3} shows the subsets $\mathcal{T}\subseteq\mathcal{S}$, and the circles $m_2''\in\mathcal{T}$ which are covered by $m_2'\in\mathcal{T}$. The proof is given in Sec.\ SC\,3 of the SM.

\subsubsection{Step 4}

Having identified which circles $m_2$ on torus $m_3$ cover all other circles $m_2'$, the final step is to cover circle $m_2$ on torus $m_3$ itself. Let $\dtilde n_2 = \tilde n_2 / e = n_2 / ed$. Looping around the circle of tori back to circle $m_2$ on torus $m_3$ takes $\lcm(\tilde n_3, \dtilde n_2)$ multiples of $(m_1, m_2, m_3)$. The first coordinate $m_1$ after $t$ loops becomes
\begin{equation}
	m_1' = m_1 (1 + t\lcm(\tilde n_3, \dtilde n_2)) \mod n_1.
\end{equation}
The possible values $m_1'$ can take are then
\begin{equation}
	m_1' \in m_1 + \langle \gcd(m_1\lcm(\tilde n_3, \dtilde n_2), n_1) \rangle_{n_1} \mod n_1.
\end{equation}
The points on the circle are thereby divided into sets $\mathcal{S}$, analogously to the result in Sec.\ \ref{sect:2d_step_2} and Fig.\ \ref{fig:2d_step_2}. The proof for this section accordingly follows directly from Secs.\ \ref{sect:2d_step_2} and SB\,2 of the SM. For each divisor $d$ of
\begin{equation}
	v = \pi_{\lcm(\tilde n_3, \dtilde n_2)}\left( \frac{n_1}{\gcd(\lcm(\tilde n_3, \dtilde n_2), n_1)} \right)
\end{equation}
an $\mathcal{S}$ exists, where the smallest non-zero difference between two elements is $d\gcd(\lcm(\tilde n_3, \dtilde n_2),n_1)$. Search then from $d$ to $d\gcd(\lcm(\tilde n_3, \dtilde n_2),n_1)$ in steps of $d$ for each $m_1'$ which satisfies
\begin{equation}
	\gcd(m_1', v) = d.
\end{equation}
Every such $m_1'$ is part of a unique $\mathcal{S}$. For each found $\mathcal{S}$, search for an $m_1 \in \mathcal{S}$ such that
\begin{equation}
	\gcd\left( m_1, \frac{n_1}{\gcd(\lcm(\tilde n_3, \dtilde n_2), n_1)} \right) = d.
\end{equation}
The resulting point $\mathbf{q} = (m_1, m_2, m_3)$ is added to $\mathcal{Q}$, if it is not already in the union of $\langle \mathbf{q}' \rangle_{(n_1,n_2,n_3)}$ for $\mathbf{q}'\in\mathcal{Q}$. This concludes the three-dimensional algorithm.

\section{The Algorithm}\label{sect:the_algorithm}

\begin{figure}[h!]
	\centering
	\includegraphics{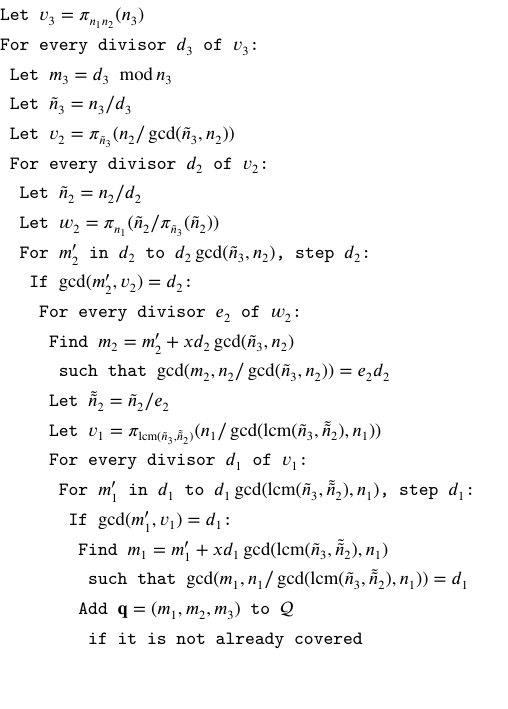}
	\vspace*{-1.3cm}
	\caption{The algorithm in pseudocode for generating $\mathcal{Q}$ on an $n_1 \times n_2 \times n_3$ grid of $\mathbf{q}$-points.}
	\label{fig:algorithm}
\end{figure}

Figure \ref{fig:algorithm} shows the algorithm for a three-dimensional $n_1\times n_2 \times n_3$ grid given with pseudocode.

The one- and two-dimensional grids are merely special cases of the three-dimensional grid, so the algorithm applies to one and two dimensions as well, with $n_2=n_3=1$ and $n_3=1$, respectively.

\section{Scaling}

\begin{figure*}
	\centering
	\includegraphics{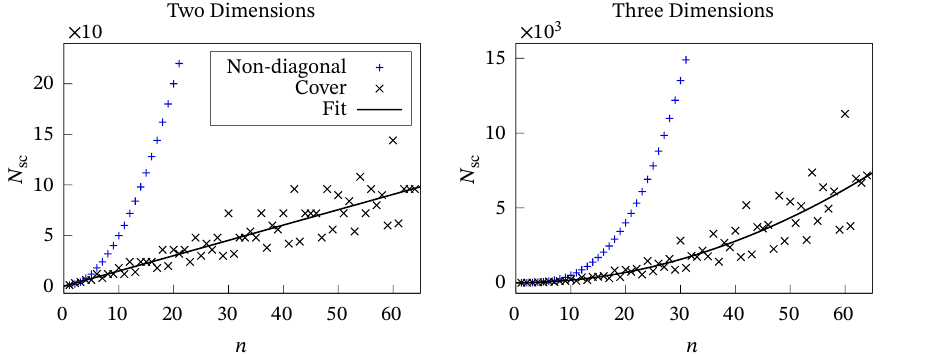}
	\caption{The number of supercells $N_\text{sc}$ to be calculated on an $n\times n$ (two dimensions) or $n\times n\times n$ (three dimensions) $\mathbf{q}$-grid, using the original non-diagonal method \cite{nondiagonal} -- approximately $n^D/2$ in $D$ dimensions -- and using the algorithm described in this work.  By fitting a function of the form $an^b$ to the data for the cover method, the obtained scalings are $\mathcal{O}(n^{1.003})$ for two dimensions and $\mathcal{O}(n^{1.999})$ for three dimensions, suggesting a scaling of $\mathcal{O}(n^{D-1})$ in $D$ dimensions. Only one supercell is needed in one dimension, in correspondence with the suggested scaling.}
	\label{fig:scaling}
\end{figure*}

On an $n\times n$ (two dimensions) or $n\times n\times n$ (three dimensions) grid of $\mathbf{q}$-points, the number of supercells $N_\text{sc}$ required by the cover method (i.e.\ the number of $\mathbf{q}$-points in $\mathcal{Q}$) is shown in Fig.\ \ref{fig:scaling} as a function of $n$. The number theoretic nature of the problem causes $N_\text{sc}$ to vary substantially with $n$, but the figure shows an overall trend that $N_\text{sc}$ scales as $\mathcal{O}(n^{D-1})$ in $D$ dimensions. Only one supercell is needed in one dimension, thereby fitting the pattern. The non-diagonal method in contrast scales as $\mathcal{O}(n^D)$, the size of the $\mathbf{q}$-grid. The diagonal method only requires a single supercell.

The values of $n$ where $N_\text{sc}$ is minimized in Fig.\ \ref{fig:scaling} are the prime numbers, where the number of supercells is given by
\begin{alignat}{2}
	N_\text{sc} &= n^2 + n + 1, \quad &&(\text{three dimensions, prime }n) \\
	N_\text{sc} &= n + 1. \quad &&(\text{two dimensions, prime }n)
\end{alignat}
This is proven in Sec.\ SD of the SM. The intuition is that the groups $\langle \mathbf{q} \rangle$ for $\mathbf{q}\in\mathcal{Q}$ overlap the least when $n$ is prime, in which case
\begin{equation}
	\langle \mathbf{q} \rangle \cap \langle \mathbf{q}' \rangle = \{\mathbf{0}\}
\end{equation}
for any two $\mathbf{q},\mathbf{q}'\in\mathcal{Q}$ with $\mathbf{q}\neq\mathbf{q}'$. Each new supercell is then commensurate with a set of $\mathbf{q}$-points which no other supercell is commensurate with, except $\mathbf{q}=\mathbf{0}$ which is commensurate with all supercells. Though prime values of $n$ in principle yield the largest computational gain, the resulting $\mathbf{q}$-grids often do not include high-symmetry points. This may be undesirable, for example when plotting phonon band structures.

With the dynamical matrices calculated through Eq.\ \eqref{eq:force}, the number of finite difference derivatives scales linearly with $N_\text{atom} N_\text{sc}$, where $N_\text{atom}$ is the number of atoms in the primitive unit cell, i.e.\ $N_\text{atom}$ derivatives in each supercell. Assuming the electronic structure calculations scale as $\left(N_\text{atom}^\text{sc}\right)^\alpha$ with the number of atoms in the supercell $N_\text{atom}^\text{sc}$, the total scaling of a frozen phonon calculation is
\begin{equation}
	\begin{split}
		&\text{(Scaling per supercell)} \\ 
		&\times \text{(Number of electronic structure calculations)} \\
		&= \left(N_\text{atom}^\text{sc}\right)^\alpha N_\text{sc}N_\text{atom}.
	\end{split}
\end{equation}
The single supercell in the diagonal supercell method contains $N_\text{atom}^\text{sc}=N_\text{atom}n^D$ atoms. The non-diagonal supercell method contains at most $N_\text{atom}^\text{sc} = N_\text{atom}n$ atoms per supercell, and so does the cover method. The theoretical scaling of the three methods is summarized in Table \ref{tab:theoretical_scaling}.

\begin{table}[h!]
	\centering
	\renewcommand{\arraystretch}{1.5}
	\begin{tabular}{c c}
		\hline
		\hline
		Diagonal & $N_\text{atom}^{\alpha+1} n^{D\alpha}$ \\
		\hline
		Non-diagonal & $N_\text{atom}^{\alpha+1} n^{D+\alpha}$ \\
		\hline
		Cover & $N_\text{atom}^{\alpha+1} n^{D-1+\alpha}$ \\
		\hline
		\hline
	\end{tabular}
	\caption{The theoretical scaling of the diagonal, non-diagonal, and cover methods in $D$ dimensions.}
	\label{tab:theoretical_scaling}
\end{table}

In the best case scenario of linear scaling electronic structure methods ($\alpha = 1$), as may theoretically be achieved with atomic orbital basis functions \cite{siesta}, the scaling of the diagonal method and the cover method are equal, while the non-diagonal is slightly worse due to the large number of supercells required. In practice $\alpha$ is often closer to three, the scaling of matrix diagonalization, thereby giving the cover method a clear advantage. It is an additional benefit of the non-diagonal and cover methods that calculations on different supercells are perfectly parallelizable.

\section{Computational Details}

The algorithm described above has been implemented in the Circles program \cite{circles} written in the Julia programming language \cite{julia}. The calculations were performed as a three-step process: 1.\ generate electronic structure input files using Circles, 2.\ perform the electronic structure calculations, and 3.\ read the calculated Hellmann-Feynman forces and summarize timings using Circles.

Finite difference derivatives were performed with a central difference derivative, requiring six calculations per atom. The finite difference displacement was set to 0.04\,Bohr in all calculations.

The non-diagonal supercell for a given $\mathbf{q}$-point (i.e.\ the smallest supercell that $\mathbf{q}$-point is commensurate with) was obtained with an algorithm implemented in Circles, which is fundamentally equivalent to the result by Lloyd-Williams and Monserrat \cite{nondiagonal}.

The electronic structure program used was SIESTA \cite{siesta} using Density Functional Theory with the Perdew-Burke-Ernzerhof \cite{pbe} exchange-correlation functional and norm-conserving pseudopotentials \cite{pseudopot} obtained from Pseudo-Dojo \cite{pseudodojo}. The electronic $\mathbf{k}$-grid was determined using a cutoff of 24\,Å \cite{kmesh}. A double-zeta basis set with polarization functions (DZP) was used \cite{basis1, basis2}. To assist convergence of electronic structure calculations on large supercells, the Self-Consistent Field (SCF) mixing weight was set to 0.05 and the mixer history to 6. These SCF settings were used for all calculations to facilitate comparison. For some calculations on CaF$_2$ with the $16\times16\times16$ grid, it was found that the generated list of $\mathbf{k}$-points caused the electronic structure calculation to be unable to converge. A cutoff of 30\,Å was used instead for these calculations.

All electronic structure calculations were performed without parallelization on an Intel Xeon Platinum 8358 CPU.

Calculations were performed on Al with a lattice parameter of $7.63$\,Bohr, CaF$_2$ with a lattice parameter of 10.42\,Bohr, and Nb$_3$Sn with a lattice parameter of 10.05\,Bohr. The lattice parameter of graphene was set to 4.65\,Bohr within the graphene plane and 20\,Bohr perpendicular to the graphene plane, to ensure adequate separation of periodic images of the two-dimensional system. The tetragonal PtS structure was calculated with a lattice parameter of 6.60\,Bohr in the $a$ and $b$ directions, and 11.56\,Bohr in the $c$ direction.

\section{Results}

\begin{figure}[h]
	\centering
	\includegraphics{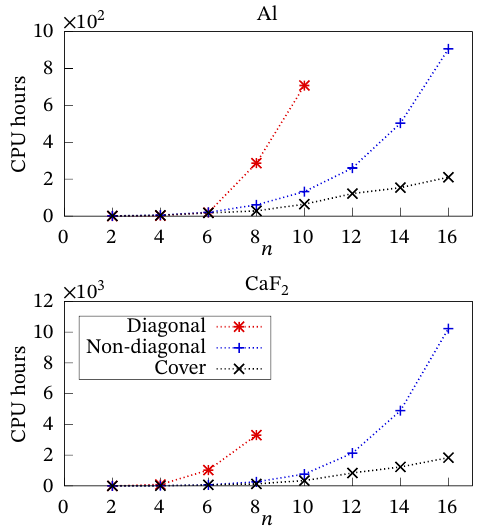}
	\caption{CPU hours for the phonon calculation of Al and CaF$_2$ on an $n\times n\times n$ grid, using the diagonal method, the non-diagonal method, and the cover method.}
	\label{fig:times}
\end{figure}

The dynamical matrices of Al and CaF$_2$ were calculated on $n\times n\times n$ grids, with $n$ ranging from 2 to 16 for the non-diagonal and cover method, 2 to 10 for Al with the diagonal method, and 2 to 8 for CaF$_2$ with the diagonal method. Figure \ref{fig:times} shows the total CPU hours for each calculation.

The diagonal method works well for coarse grids ($n=2,4$), but rapidly increases in CPU hours as the grid becomes denser. As $n$ increases above $8$, the non-diagonal method starts to increase significantly, while the cover method stays relatively low for all grids. The cover method is approximately five times faster than the non-diagonal method for $n=16$. The figure suggests that this difference should only increase for denser grids.

\begin{figure}
	\centering
	\includegraphics{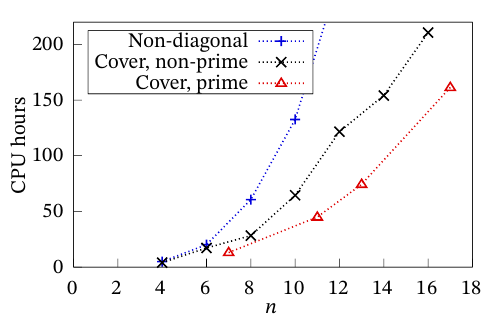}
	\caption{CPU hours for the Al phonon calculation on an $n\times n\times n$ grid using the cover method for prime and non-prime grids, as well as the non-diagonal method.}
	\label{fig:times_prime}
\end{figure}

The computational gain is expected to be highest for prime values of $n$. Figure \ref{fig:times_prime} shows the CPU hours spent calculating the phonons of Al with the cover method using prime and non-prime grids.

The prime grids show the expected trend. Note in particular that for prime $n$, the calculation is in fact faster than the calculation for $n-1$ across the board.

The small number of calculations required for the diagonal method can be an advantage over the two other methods for very coarse grids. The $n=2$ and $n=4$ grids of Al are faster with the diagonal method than the cover method, and comparable for $n=6$. The phonon calculation of CaF$_2$ with the diagonal method is faster for $n=2$, but significantly slower than the other two for $n=4$ and above. The complete timings are given in the SM, Sec.\ SE. This effect is caused by the small unit cells of both Al and CaF$_2$, which contain one and three atoms in the primitive unit cell, respectively. For the Nb$_3$Sn system with eight atoms per unit cell, the cover method is almost twice as fast as the diagonal method for a phonon calculation on an $n \times n \times n$ grid with $n=2$, and ten times faster for $n=4$.

\begin{figure}
	\centering
	\includegraphics{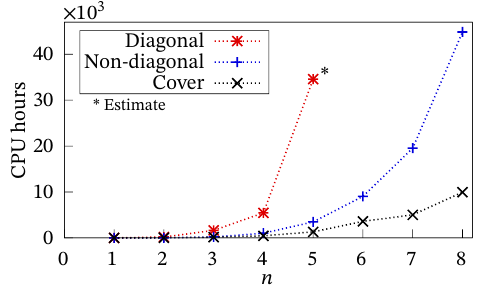}
	\caption{CPU hours for the PtS phonon calculation on a $2n\times 2n\times n$ grid using the diagonal, non-diagonal, and cover method. The diagonal $n=5$ calculation is estimated from the average time of the first SCF iteration of the $n=5$ calculations, and the average number of SCF iterations of the $n=4$ calculation.}
	\label{fig:times_PtS}
\end{figure}

While all three systems discussed above are cubic, the PtS system is tetragonal and about twice as long in the $c$ direction as in the $a$ and $b$ directions. A grid of $2n\times 2n\times n$ is therefore more appropriate for this system. Figure \ref{fig:times_PtS} shows the CPU hours for phonon calculations on such grids for $n=1$ to $8$ for the non-diagonal and cover methods. Calculations for the diagonal method were performed for $n=1$ to $4$, but limitations on the employed cluster meant that the $n=5$ calculation could not be completed. It is instead estimated from the average time of the first SCF iteration of the $n=5$ calculations, multiplied by the number of SCF calculations (24) and the average number of SCF iterations on the $n=4$ calculation (16). This is expected to slightly underestimate the total computational time, as it does not include processes performed before the first iteration (e.g.\ calculating the overlap matrix) and after the last iteration (e.g.\ calculating the forces).

The result follows the expected trend. The diagonal method increases quickly in CPU hours. The non-diagonal method is good on modest grids, but becomes expensive for denser grids, while the cover method increases the least in computational cost.

\begin{figure}
	\centering
	\includegraphics{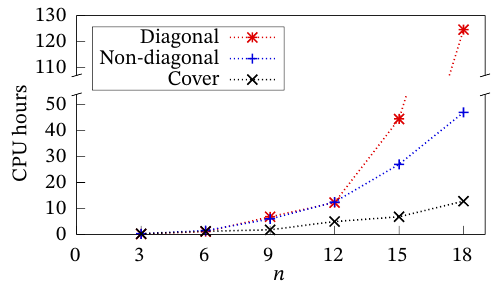}
	\caption{CPU hours for the graphene phonon calculation on an $n\times n$ grid using the diagonal, non-diagonal, and cover method. Note the broken $y$-axis.}
	\label{fig:times_graphene}
\end{figure}

Figure \ref{fig:times_graphene} shows the CPU hours for phonon calculations on the two-dimensional graphene system, calculated on $n\times n$ grids.
The diagonal supercell of a two-dimensional grid remains relatively small, and the diagonal and non-diagonal methods accordingly show very similar timings up to $n=12$, after which the diagonal method increases significantly above the non-diagonal method. The cover method yields a computational speed-up comparable to the results of the three-dimensional systems. 

\section{Conclusion}

This work has presented a method to significantly speed up frozen phonon calculations on dense $\mathbf{q}$-grids. The cover method is an extension of the non-diagonal supercell method, obtaining the minimal number of supercells which need to be calculated, while still covering every point on the chosen $\mathbf{q}$-grid. The method has been proven to be the optimal solution to the problem. Numerical results have shown the computational advantage of the method for a variety of systems, based on an open source implementation of the method in the Circles program \cite{circles}.

Symmetry has been omitted from this study for the sake of simplicity, but would be a natural way to reduce the number of calculations even further. If $\mathbf{q}$ and $\mathbf{q}'$ are related by some symmetry element $g$, then every element in $\langle \mathbf{q}' \rangle$ is related to an element in $\langle \mathbf{q} \rangle$ through $g$, implying that each $\mathbf{D}(\mathbf{p})$ for $\mathbf{p}\in\langle\mathbf{q}'\rangle$ can be obtained from some $\mathbf{D}(\mathbf{p})$ for $\mathbf{p}\in\langle\mathbf{q}\rangle$ by symmetry. That being said, the computational advantage of the cover method is completely independent of the system, and provides the same efficiency on highly symmetric as on highly asymmetric structures.

Though only applied to the calculation of dynamical matrices here, the method is equally suitable for any other property relying on the frozen phonon approach. Electron-phonon coupling and higher order force constants, for example, are ample candidates for further application of the method.

\begin{acknowledgments}
I am grateful to Frank Jensen for fruitful discussions and for his help revising this manuscript. The numerical results presented in this work were obtained at the Center for Scientific Computing, Aarhus.
\end{acknowledgments}

\bibliography{main}

\end{document}


\title{Supplemental Material: Frozen Phonons on Dense \textbf{q}-Grids Using Minimal Sets of Non-Diagonal Supercells}

\author{Sebastian Riis Thomsen}
\email[]{seb@chem.au.dk}
\affiliation{Department of Chemistry, Aarhus University, Denmark}
\date{\today}

\maketitle

\section{Properties of the Greatest Common Divisor}\label{sect:properties_gcd}

In the following, $a$, $b$, and $c$ will always denote integers. By the fundamental theorem of arithmetic \cite{numbertheory} $a$, $b$, and $c$ can be written in terms of their \textit{prime factorization}
\begin{equation}
	a = \prod_p p^{a_p}
\end{equation}
for non-negative integers $a_p$, where the product is over all primes, and likewise for $b$ and $c$. This is often very helpful when dealing with greatest common divisors and least common multiples, since
\begin{equation}
	\gcd(a, b) = \prod_p p^{\min(a_p, b_p)} \quad \text{and} \quad \lcm(a, b) = \prod_p p^{\max(a_p, b_p)}.
\end{equation}

\subsection{Property \the\value{subsection}}\label{sect:b_begone}

\noindent
Assume $\gcd(b, c)=1$. Then $\gcd(ab,c) = \gcd(a,c)$.

\textbf{Proof:}

\noindent
For a given prime $p$, $\gcd(b, c)=1$ implies that either $b_p=0$ or $c_p=0$. If $b_p=0$ then
\begin{equation}
	\min(a_p+b_p, c_p) = \min(a_p, c_p)
\end{equation}
and if $c_p=0$ then
\begin{equation}
	\min(a_p+b_p, c_p) = 0 = \min(a_p, c_p).
\end{equation}
In either case $\min(a_p+b_p, c_p) = \min(a_p, c_p)$, so
\begin{equation}
	\gcd(ab, c) = \prod_p p^{\min(a_p+b_p, c_p)} = \prod_p p^{\min(a_p, c_p)} = \gcd(a, c).
\end{equation}
{\hfill$\blacksquare$}

\subsection{Property \the\value{subsection}}\label{sect:c_begone}

\noindent
Assume $\gcd(a, c) = 1$ and $c\mid b$. Then $\gcd(a, b/c) = \gcd(a, b)$.

\textbf{Proof:}

\noindent
Let $b'=b/c$. Then, by the property in Sec.\ \ref{sect:b_begone}, 
\begin{equation}
	\gcd(a, b) = \gcd(a, b'c) = \gcd(a, b') = \gcd(a, b/c).
\end{equation}
{\hfill$\blacksquare$}

\subsection{Property \the\value{subsection}}\label{sect:divide_into_gcd}

\noindent
Assume $c\mid a$ and $c\mid b$. Then $\gcd(a, b)/c = \gcd(a/c, b/c)$.

\textbf{Proof:}

\noindent
\begin{equation}
	\frac{\gcd(a, b)}{c} = \prod_p p^{\min(a_p, b_p) - c_p} = \prod_p p^{\min(a_p - c_p, b_p - c_p)} = \gcd\left( \frac{a}{c}, \frac{b}{c} \right).
\end{equation}
{\hfill$\blacksquare$}

\subsection{Property \the\value{subsection}}\label{sect:bc_split}

\noindent
Assume $\gcd(b, c) = 1$. Then $\gcd(a, bc) = \gcd(a, b)\gcd(a, c)$.

\textbf{Proof:}

\noindent
For a given prime $p$, $\gcd(b,c)=1$ implies that either $b_p=0$, so
\begin{equation}
	\min(a_p, b_p+c_p) = \min(a_p, c_p) = \min(a_p, b_p) + \min(a_p, c_p),
\end{equation}
or $c_p=0$, so
\begin{equation}
	\min(a_p, b_p+c_p) = \min(a_p, b_p) = \min(a_p, b_p) + \min(a_p, c_p).
\end{equation}
In either case
\begin{equation}
	\begin{split}
		\gcd(a, bc) &= \prod_p p^{\min(a_p, b_p+c_p)} \\
			&= \prod_p p^{\min(a_p, b_p) + \min(a_p, c_p)} \\
			&= \left( \prod_p p^{\min(a_p, b_p)} \right)\left( \prod_p p^{\min(a_p, c_p)} \right) \\
			&= \gcd(a, b)\gcd(a, c).
	\end{split}
\end{equation}
{\hfill$\blacksquare$}

\subsection{Property \the\value{subsection}}\label{sect:gcd_lcm}

\noindent
$\gcd(a, b) \lcm(a, b) = ab$.

\textbf{Proof:}

\noindent
For any two integers $c$ and $d$,
\begin{equation}
	\max(c, d) + \min(c, d) = c + d.
\end{equation}
From this it is found that
\begin{equation}
	\begin{split}
		\gcd(a, b) \lcm(a, b) &= \left( \prod_p p^{\min(a_p, b_p)} \right) \left( \prod_p p^{\max(a_p, b_p)} \right) \\
			&= \left( \prod_p p^{\min(a_p, b_p) + \max(a_p, b_p)} \right) \\
			&= \left( \prod_p p^{a_p + b_p} \right) \\
			&= \left( \prod_p p^{a_p} \right)\left( \prod_p p^{b_p} \right) 
			= ab.
	\end{split}
\end{equation}
{\hfill$\blacksquare$}

\subsection{Property \the\value{subsection}}\label{sect:gcd_set_of_ints}

\noindent
Let $\{b_i\}$ be a set of integers. Then,
\begin{equation}
	\gcd\left(a, \prod_i b_i\right) = 1 \quad \Longleftrightarrow \quad \forall i: \gcd(a, b_i)=1.
\end{equation}

\textbf{Proof:}

\noindent
Using the prime factorizations,
\begin{equation}
	\gcd\left(a, \prod_i b_i\right) = \prod_p p^{\min\left(a_p, \sum_i b_{ip}\right)} = 1
\end{equation}
implies that for any prime $p$ either $a_p=0$ or $\sum_i b_{ip}=0$. Since the exponents must be non-negative, the sum equals zero if and only if $b_{ip}=0$ for every $i$. This in turn means that for all $i$ and all $p$, $\min(a_p, b_{ip})=0$, so
\begin{equation}
	\forall i: \gcd(a, b_i) = \prod_p p^{\min(a_p, b_{ip})} = \prod_p p^0 = 1.
\end{equation}
{\hfill$\blacksquare$}

\subsection{Property \the\value{subsection}}\label{sect:mod_b_b}

\noindent
$\gcd(a+b, b) = \gcd(a, b)$.

\textbf{Proof:}

\noindent
If $d\mid a$ and $d\mid b$ then $d\mid a+b$. If conversely $d\mid b$ and $d\mid a+b$ then $d\mid a$. This implies that the common divisors of $a$ and $b$ are the same as the common divisors of $b$ and $a+b$. The greatest common divisor must then also be the same.
{\hfill$\blacksquare$}

\subsection{Property \the\value{subsection}}\label{sect:c_divides_ab}

\noindent
Assume $c\mid ab$ and $\gcd(b,c)=1$. Then $c\mid a$.

\textbf{Proof:}

\noindent
The condition $c\mid ab$ implies that $a_p+b_p-c_p\ge 0$ for every prime $p$. For those primes $p$ where $p\mid c$, $\gcd(b,c)=1$ implies $p\nmid b$, so $0\le a_p+b_p-c_p = a_p-c_p$. For those primes $p$ where $p\nmid c$, $c_p=0$, so $a_p-c_p=a_p \ge 0$. Then $a_p-c_p \ge 0$ for every $p$, so $c\mid a$.
{\hfill$\blacksquare$}

\subsection{Property \the\value{subsection}}\label{sect:lcm_div}

\noindent
Assume $\gcd(a, c) = 1$ and $c \mid b$. Then $\lcm(a,b)/c = \lcm(a, b/c)$.

\textbf{Proof:}

\noindent
For any prime $p$ which divides $a$, $c_p=0$, so
\begin{equation}
	\max(a_p,b_p)-c_p = \max(a_p,b_p) = \max(a_p,b_p-c_p).
\end{equation}
For any prime $p$ which conversely divides $c$, $a_p=0$, so
\begin{equation}
	\max(a_p,b_p)-c_p = b_p-c_p = \max(a_p,b_p-c_p).
\end{equation}
As a result,
\begin{equation}
	\frac{\lcm\left(a,b\right)}{c} = \prod_p p^{\max(a_p,b_p)-c_p} = \prod_p p^{\max(a_p,b_p-c_p)} = \lcm\left(a,\frac{b}{c}\right).
\end{equation}
{\hfill$\blacksquare$}


\subsection{Property \the\value{subsection}}\label{sect:smallest_integer}

\noindent
Assume $b\nmid a$. The smallest non-zero element in the group $\langle a \rangle_b$ is $\gcd(a, b)$. (When $b\mid a$, $\langle a \rangle_b = \{0\}$)

\textbf{Proof:}

\noindent
The group $\langle a \rangle_b$ consists of those integers $a'$ with $0\le a' < b$, where $a'=ax+by$ for some integers $x$ and $y$. Then $\gcd(a,b)\mid a,b$ implies $\gcd(a,b)\mid ax+by=a'$. By B\'ezout's identity \cite{numbertheory}, two integers $x$ and $y$ exist, such that $ax+by=\gcd(a,b)$. The assumption $b \nmid a$ implies $\gcd(a,b)<b$, so $\gcd(a,b)\in\langle a\rangle_b$. The smallest non-zero, positive integer divisible by $\gcd(a,b)$ is (of course) $\gcd(a,b)$ itself, and since every $a' \in \langle a \rangle_b$ is divisible by $\gcd(a,b)$, $\gcd(a,b)$ must be the smallest non-zero integer in $\langle a \rangle_b$.
{\hfill$\blacksquare$}

\subsection{Property \the\value{subsection}}\label{sect:t_exists}

\noindent
Let $k$ be an integer. An integer $t$ exists such that $\gcd(k+tb, a) = \gcd(k,b,a)$.

\textbf{Proof:}

\noindent
Note first that $\gcd(k,b,a)\mid k$ and $\gcd(k,b,a)\mid b$ implies $\gcd(k,b,a)\mid k+tb$ for any $t$. Then
\begin{equation}
	\begin{split}
		\gcd(k, b, a) \mid \gcd(k+tb, a) \quad \Longrightarrow \quad &\frac{\gcd(k+tb, a)}{\gcd(k,b,a)} \\
			&= \gcd\left( \frac{k}{\gcd(k,b,a)} + t\frac{b}{\gcd(k,b,a)}, \frac{a}{\gcd(k,b,a)} \right) \\
			&\equiv \gcd(k'+tb', a').
	\end{split}
\end{equation}
It needs to be shown that
\begin{equation}\label{eq:kba_left_right}
	\gcd(k', b', a') = 1 \quad \Longleftrightarrow \quad \gcd(k'+tb', a') = 1.
\end{equation}
Assume first that $\gcd(k'+tb', a') = 1$. Then
\begin{equation}
	\gcd(k',b',a') \mid \gcd(k'+tb', a') = 1 \quad \Longrightarrow \quad \gcd(k',b',a') = 1.
\end{equation}
This proves the $\Leftarrow$ in Eq. \eqref{eq:kba_left_right}.

Assume then that $\gcd(k',b',a') = 1$. For every prime $p$ which divides $a'$,
\begin{equation}
	\gcd(k', b', a') = 1 \quad \Longleftrightarrow \quad \forall p \mid a': \gcd(k', b', p) = 1.
\end{equation}
Consider first those $p$ which also divide $b'$. Then
\begin{equation}
	\gcd(k', b', a') = 1 \quad \Longrightarrow \quad \gcd(k', p) = 1
\end{equation}
and
\begin{equation}
	\gcd(k'+tb', p) = \gcd(k', p) = 1.
\end{equation}
The property $\gcd(a+b, b) = \gcd(a, b)$ has been used (Sec.\ \ref{sect:mod_b_b}). Now consider those $p$ which do not divide $b'$. For any integer $n$ and prime $p$, $\gcd(n,p) =$ either $p$ or $1$. In the former case, $p\mid n$. Then \cite{numbertheory}
\begin{equation}
	\gcd(b', p) = 1 \quad \Longrightarrow \quad \exists b'^{-1}\in\mathbb{Z}: b'^{-1} b' = 1 \mod p.
\end{equation}
For any $t'_p$, $\gcd(k'+t'_pb', p)$ is either $p$ or $1$. In the former case,
\begin{equation}
	k'+t'_pb' = 0 \mod p \quad \Longrightarrow \quad t'_p = -k'b'^{-1} \mod p.
\end{equation}
This $t'_p$ is unique $\mathrm{mod}~p$, so choosing any other $t_p\neq t'_p$ yields $\gcd(k'+t_pb', p)=1$. This provides a set of equations
\begin{equation}
	\forall p\mid a', p\nmid b': t = t_p \smod p.
\end{equation}
By the Chinese Remainder Theorem \cite{numbertheory}, there exists a $t$ such that all equations are solved simultaneously. Finally,
\begin{equation}
	\begin{split}
		\forall p\mid a': \gcd(k'+t b', p) = 1 \quad &\Longrightarrow \quad \gcd(k'+tb', a')=1 \\
			&\Longrightarrow \quad \gcd(k+tb, a) = \gcd(k,b,a).
	\end{split}
\end{equation}
{\hfill$\blacksquare$}

\subsection{Property \the\value{subsection}}\label{sect:pi_divisors}

\noindent
Any prime $p$ which divides $\pi_b(a)$ also divides $b$.

\textbf{Proof:}

\noindent
Write $a=\pi_b(a)r$ where (by definition) $r$ is the largest divisor of $a$ coprime to $b$. The proof proceeds by contradiction.

Let $p$ be a prime which divides $\pi_b(a)$, and assume that it does not divide $b$. Then $\gcd(b, p)=1$ and, by the property in Sec.\ \ref{sect:b_begone},
\begin{equation}
	\gcd(b, pr) = \gcd(b, r) = 1.
\end{equation}
Since $p$ divides $\pi_b(a)$, $pr$ must divide $a$. Then $pr$ is a divisor of $a$ coprime to $b$. This is a contradiction, since $pr>r$, whereby the assumption that $p\nmid b$ cannot hold.
{\hfill$\blacksquare$}

\subsection{Property \the\value{subsection}}\label{sect:pi_coprime_coprime}

\noindent
Any integer coprime to $b$ is also coprime to $\pi_b(a)$.

\textbf{Proof:}

\noindent
Assume $\gcd(c, b)=1$. Then
\begin{equation}
	\forall p\mid b: \gcd(c, p) = 1.
\end{equation}
Any prime $p$ which divides $\pi_b(a)$ also divides $b$ (Sec.\ \ref{sect:pi_divisors}), so the above implies
\begin{equation}
	\forall p\mid \pi_b(a): \gcd(c, p) = 1 \quad \Longrightarrow \quad \gcd(c, \pi_b(a)) = 1.
\end{equation}
{\hfill$\blacksquare$}

\subsection{Property \the\value{subsection}}\label{sect:pia_pib}

\noindent
Assume $a\mid b$. Then $\pi_c(a) \mid \pi_c(b)$.

\textbf{Proof:}

\noindent
Since $a \mid b$, there exists an integer $d$ such that 
\begin{equation}
	b=da=\left(\pi_c(d)r_d\right)\left(\pi_c(a)r_a\right) = \left(\pi_c(d)\pi_c(a)\right)\left(r_dr_a\right), \qquad \gcd(c,r_d)=\gcd(c,r_a)=1.
\end{equation}
Writing $b=\pi_c(b)r_b$ it follows that $r_dr_a \mid r_b$, since $r_b$ is the largest divisor of $b$ coprime to $c$, and $r_dr_a$ is certainly a divisor of $b$ coprime to $c$. Then $r_b = er_d r_a$ and $\pi_c(b) = \pi_c(d) \pi_c(a) / e$ for some integer $e$. However, $\gcd(e,c)=1$ implies $\gcd(e,\pi_c(d)\pi_c(a))=1~\Rightarrow~e=1$, and $\pi_c(b) = \pi_c(d) \pi_c(a)~\Rightarrow~\pi_c(a) \mid \pi_c(b)$.

{\hfill$\blacksquare$}

\subsection{Property \the\value{subsection}}\label{sect:possible_values}

\noindent
Let $x$ be an integer. The values taken by $\gcd(1+xb, a)$ for all $x$ are the divisors of $a/\pi_b(a)$, i.e.\
\begin{equation}
	\{ \gcd(1+xb, a) : x\in\mathbb{Z} \} =~\text{divisors of}~\frac{a}{\pi_b(a)}.
\end{equation}

\clearpage

\textbf{Proof:}

\noindent
First note that $\gcd(1+xb, b)=\gcd(1,b)=1$ (see Sec.\ \ref{sect:mod_b_b}), so for any integer $n$,
\begin{equation}
	\begin{split}
		\gcd(1+xb, b^n) = \gcd(1+xb, bb^{n-1}) &= \gcd(1+xb, b^{n-1}) \\
			&= \gcd(1+xb, b^{n-2}) \\
			&= \dots \\
			&= \gcd(1+xb, b) = 1.
	\end{split}
\end{equation}
Then $\gcd(1+xb, \gcd(a, b^n))=1$ for any $n$, so
\begin{equation}
	\gcd(1+xb, a) = \gcd\left( 1+xb, \frac{a}{\gcd(a, b^n)} \right).
\end{equation}
Let $n$ be an integer large enough that
\begin{equation}
	\gcd\left( b, \frac{a}{\gcd(a, b^n)} \right) = 1.
\end{equation}
Such an integer always exists, for example by choosing $n$ such that $nb_p \ge a_p$ for every $p\mid b$, as shown below. Then
\begin{equation}
	\langle b \rangle_{a/\gcd(a, b^n)} = \mathbb{Z}_{a/\gcd(a, b^n)} \quad
	\Longrightarrow \quad \left\{1 + bx \smod \frac{a}{\gcd(a, b^n)} : x \in \mathbb{Z}\right\} = \mathbb{Z}_{a/\gcd(a, b^n)}.
\end{equation}
This implies that
\begin{equation}
	\begin{split}
	\left\{\gcd\left(1 + bx, \frac{a}{\gcd(a, b^n)}\right) : x \in \mathbb{Z}\right\}
	&= \left\{ \gcd\left(x, \frac{a}{\gcd(a, b^n)}\right) : x \in \mathbb{Z} \right\} \\
	&= \text{ divisors of } \frac{a}{\gcd(a, b^n)}.
	\end{split}
\end{equation}

It then needs to be shown that $\pi_b(a) = \gcd(a, b^n)$. Since
\begin{equation}
	a = \gcd(a, b^n) \frac{a}{\gcd(a, b^n)}
\end{equation}
this is equivalent to showing that $a/\gcd(a, b^n)$ is the largest divisor of $a$ coprime to $b$. It is certainly a divisor of $a$, and by definition of $n$ also coprime to $b$. The prime factorization is
\begin{equation}
	\frac{a}{\gcd(a, b^n)} = \prod_p p^{a_p - \min(a_p, nb_p)}
\end{equation}
so
\begin{equation}
	\begin{split}
		\gcd\left( b, \frac{a}{\gcd(a, b^n)} \right) &= \prod_p p^{\min(b_p, a_p - \min(a_p, nb_p))} = 1 \\
			&\Longrightarrow\quad \forall p: \min(b_p, a_p - \min(a_p, nb_p)) = 0.
	\end{split}
\end{equation}
This implies that either $b_p = 0$, in which case $p \nmid b$, or $a_p - \min(a_p, nb_p)=0$, implying $a_p \le nb_p$. Then $a/\gcd(a, b^n)$ can be written as
\begin{equation}
	\begin{split}
		\frac{a}{\gcd(a, b^n)} &= \left( \prod_{p \nmid b} p^{a_p - \min(a_p, nb_p)} \right) \left( \prod_{p \mid b} p^{a_p - \min(a_p, nb_p)} \right) \\
			&= \left( \prod_{p \nmid b} p^{a_p} \right) \left( \prod_{p \mid b} p^0 \right) \\
			&= \prod_{p \nmid b} p^{a_p}
	\end{split}
\end{equation}
which is clearly the largest divisor of $a$ coprime to $b$, implying $\gcd(a, b^n)=\pi_{b}(a)$. 
{\hfill$\blacksquare$}

\section{Proofs for the Two-Dimensional Algorithm}

\subsection{Step 1}\label{sect:proof_2d_step_1}

Start at the point $(m_1, m_2)$. Looping around the torus $t$ times, and then adding $k$ multiples of $(m_1, m_2)$, the starting point becomes $(m_1', m_2')$ given by
\begin{equation}
	m_2' = m_2 \left(k+t\frac{n_2}{\gcd(m_2, n_2)}\right) = m_2 k + t\frac{m_2}{\gcd(m_2, n_2)} n_2 = m_2 k \mod n_2
\end{equation}
and
\begin{equation}\label{eq:mprime_=_m}
	m_1' = m_1 \left(k+t\frac{n_2}{\gcd(m_2, n_2)}\right) \mod n_1.
\end{equation}
Assuming circle $m_2$ is covered, there exists some value of $t$ and some $m_1\in\mathbb{Z}_{n_1}$ such that $(m_1', m_2')$ is covered by $(m_1, m_2)$ for
\begin{equation}\label{eq:mprime_in_k+tn}
	m_1' \in \left\langle k+t\frac{n_2}{\gcd(m_2, n_2)} \right\rangle_{n_1} = \left\langle \gcd\left(k+t\frac{n_2}{\gcd(m_2, n_2)}, n_1\right) \right\rangle_{n_1}.
\end{equation}
It is shown in Sec.\ \ref{sect:t_exists} that a value $t'$ exists such that
\begin{equation}
	\gcd\left(k+t'\frac{n_2}{\gcd(m_2, n_2)}, n_1\right) = \gcd\left(k,\frac{n_2}{\gcd(m_2, n_2)}, n_1\right).
\end{equation}
Then $(m_1',m_2')$ is covered by $(m_1, m_2)$ for 
\begin{equation}\label{eq:mprime_in_kn}
	m_1' \in \left\langle \gcd\left(k,\frac{n_2}{\gcd(m_2, n_2)}, n_1\right) \right\rangle_{n_1}.
\end{equation}
The $\gcd$ in Eq.\ \eqref{eq:mprime_in_kn} divides the $\gcd$ on the right-hand side of Eq.\ \eqref{eq:mprime_in_k+tn}, whereby the generated group in Eq.\ \eqref{eq:mprime_in_k+tn} is a subgroup of the generated group in Eq. \eqref{eq:mprime_in_kn} for every $t$. In other words, for every value of $t$, each $m_1'$ in Eq.\ \eqref{eq:mprime_=_m} is also in the generated group in Eq. \eqref{eq:mprime_in_kn}. It is concluded that circle $m_2'$ is covered by circle $m_2$ if and only if
\begin{equation}\label{eq:cover_condition_si}
	\gcd\left(k,\frac{n_2}{\gcd(m_2, n_2)}, n_1\right) = 1.
\end{equation}
This is the same as Eq.\ (21) in the main text.

It is now shown that any $m_2$ which does \textit{not} divide $\pi_{n_1}(n_2)$ covers $m_2'$ only if the divisor 
\begin{equation}
	d=\gcd(m_2, \pi_{n_1}(n_2))
\end{equation}
covers $m_2'$. First,
\begin{equation}
	\gcd(m_2,n_2) = \gcd(m_2, \pi_{n_1}(n_2)) \gcd(m_2, r) = d\gcd(m_2, r).
\end{equation}
Both $m_2$ and $d$ reach $m_2'$ by assumption, whereby $m_2' = dk' = m_2k \smod n_2$. It is seen that $d\mid m_2$ by definition, so $k'=km_2/d \smod n_2/d$. Following from this,
\begin{equation}
	\begin{split}
	\gcd\left( k', \frac{n_2}{d} \right) &= \gcd\left( k\frac{m_2}{d}, \frac{n_2}{d} \right) \\
		&= \gcd\left( k\frac{m_2}{d}, \frac{\pi_{n_1}(n_2)}{d} \right) \gcd\left( k\frac{m_2}{d}, r \right) \\
		&= \gcd\left( k, \frac{\pi_{n_1}(n_2)}{d} \right) \gcd\left( k\frac{m_2}{d}, r \right).
	\end{split}
\end{equation}
The last equality follows from $\gcd(b,c)=1\Rightarrow\gcd(ab,c)=\gcd(a,c)$ (Sec.\ \ref{sect:b_begone}) and 
\begin{equation}
	\gcd(m_2,\pi_{n_1}(n_2))=d \quad \Longrightarrow \quad \gcd\left(\frac{m_2}{d}, \frac{\pi_{n_1}(n_2)}{d}\right) = 1,
\end{equation}
(section \ref{sect:divide_into_gcd}). Inserting $m_2'=km_2\smod n_2$ into Eq.\ \eqref{eq:cover_condition_si},
\begin{equation}\label{eq:km_condition}
	\begin{split}
		\gcd\left( k, \frac{n_2}{\gcd(m_2, n_2)}, n_1 \right) &= \gcd\left( k, \frac{n_2}{d\gcd(m_2, r)}, n_1 \right) \\
			&= \gcd\left( k, \frac{n_2}{d}, n_1 \right) \\
			&= \gcd\left( k, \frac{\pi_{n_1}(n_2)r}{d}, n_1 \right) \\
			&= \gcd\left( k, \frac{\pi_{n_1}(n_2)}{d}, n_1 \right)
	\end{split}
\end{equation}
using that $\gcd(a,c)=1$ and $c\mid b$ implies $\gcd(a,b/c)=\gcd(a,b)$ (Sec.\ \ref{sect:c_begone}). Inserting $m_2'=k'd$ on the other hand,
\begin{equation}
	\begin{split}
		\gcd\left(k', \frac{n_2}{\gcd(d, n_2)}, n_1 \right) = \gcd\left(k', \frac{n_2}{d}, n_1 \right) &= \gcd\left( \gcd\left( k, \frac{\pi_{n_1(n_2)}}{d} \right) \gcd\left( k\frac{m_2}{d}, r \right), n_1 \right) \\
			&= \gcd\left( k, \frac{\pi_{n_1}(n_2)}{d}, n_1 \right).
	\end{split}
\end{equation}
This is the same relation as Eq.\ \eqref{eq:km_condition}, so $m_2$ covers $m_2'$ (i.e.\ fulfills Eq.\ \eqref{eq:cover_condition_si}) if and only if $d$ covers $m_2'$. It is in conclusion unnecessary to include a $\mathbf{q}$-point in $\mathcal{Q}$ where $m_2$ is not a divisor of $\pi_{n_1}(n_2)$, since such an $m_2$ covers the same circles as the divisor $d=\gcd(m_2, \pi_{n_1}(n_2))$ of $\pi_{n_1}(n_2)$.

Next it is shown that every divisor of $\pi_{n_1}(n_2)$ must be included in $\mathcal{Q}$. For a given $m_2'\in\mathbb{Z}_{n_2}$, let
\begin{equation}\label{eq:m2_divisor}
	m_2 = \gcd(m_2', \pi_{n_1}(n_2)) \quad \Longrightarrow \quad \gcd\left( \frac{m_2'}{m_2}, \frac{\pi_{n_1}(n_2)}{m_2} \right) = 1.
\end{equation}
Since $m_2\mid \pi_{n_1}(n_2)$ it also divides $n_2$, so $\gcd(m_2, n_2) = m_2$. Furthermore, writing $m_2'$ as $m_2'=m_2k$,
\begin{equation}
	\gcd\left( k, \frac{n_2}{m_2} \right) = \gcd\left( \frac{m_2'}{m_2}, \frac{\pi_{n_1}(n_2) r}{m_2} \right) = \gcd\left( \frac{m_2'}{m_2}, r \right) = \gcd(k, r) \quad \text{with }\gcd(r, n_1) = 1.
\end{equation}
The relation $\gcd(a,b)=1\Rightarrow\gcd(a,bc)=\gcd(a,c)$ (see Sec.\ \ref{sect:b_begone}) has been used in the middle equality. Inserting $k$ into Eq.\ \eqref{eq:cover_condition_si},
\begin{equation}
	\gcd\left( k, \frac{n_2}{\gcd(m_2, n_2)}, n_1 \right) = \gcd\left( k, \frac{n_2}{m_2}, n_1 \right) = \gcd(k, r, n_1) = \gcd(k, \gcd(r, n_1)) = 1.
\end{equation}
Then $m_1'$ can take any value in $\mathbb{Z}_{n_1}$ by Eq.\ \eqref{eq:mprime_in_kn}, so circle $m_2'$ is completely covered by circle $m_2$.

Now consider some other divisor $m_2''$ of $\pi_{n_1}(n_2)$ which also reaches $m_2'$, i.e.\ $m_2'=m_2''k'$, but with
\begin{equation}
	m_2'' < \gcd(m_2', \pi_{n_1}(n_2)) \quad \Longrightarrow \quad \gcd\left( \frac{m_2'}{m_2''}, \frac{\pi_{n_1}(n_2)}{m_2''} \right) = \gcd\left( k', \frac{\pi_{n_1}(n_2)}{m_2''} \right) > 1.
\end{equation}
Using the relation $\gcd(b,c)=1\Rightarrow\gcd(a,bc)=\gcd(a,b)\gcd(a,c)$ (Sec.\ \ref{sect:bc_split}), 
\begin{equation}
	\gcd\left( k', \frac{n_2}{m_2''} \right) = \gcd\left( k', \frac{\pi_{n_1}(n_2) r}{m_2''} \right) = \gcd\left( k', \frac{\pi_{n_1}(n_2)}{m_2''} \right) \gcd(k', r) > \gcd(k', r).
\end{equation}
Since $\gcd(k', \pi_{n_1}(n_2)/m_2'')$ is a divisor of $\pi_{n_1}(n_2)/m_2''$, which in turn is a divisor of $\pi_{n_1}(n_2)$, it cannot be coprime to $\pi_{n_1}(n_2)$, and thereby not coprime to $n_1$ either. This follows from Sec.\ \ref{sect:pi_coprime_coprime}. Because of this,
\begin{equation}
	\begin{split}
		\gcd\left( k', \frac{n_2}{\gcd(m_2'', n_2)}, n_1 \right) = \gcd\left( k', \frac{n_2}{m_2''}, n_1 \right) &= \gcd\left( k', \frac{\pi_{n_1}(n_2)}{m_2''}, n_1 \right) \gcd(k', r, n_1) \\
			&= \gcd\left( k', \frac{\pi_{n_1}(n_2)}{m_2''}, n_1 \right) > 1. \\
	\end{split}
\end{equation}
Any such $m_2''$ does therefore not cover $m_2'$. The only divisor of $\pi_{n_1}(n_2)$ which covers $m_2'$ is $m_2$, as it is given by Eq.\ \eqref{eq:m2_divisor}. For every divisor $m_2$ of $\pi_{n_1}(n_2)$, there exists at least one $m_2'$ fulfilling Eq.\ \eqref{eq:m2_divisor} -- a notable example is $m_2'=m_2$. Every divisor of $\pi_{n_1}(n_2)$ must in conclusion be the second coordinate of at least one $\mathbf{q}\in\mathcal{Q}$.

\subsection{Step 2}\label{sect:proof_2d_step_2}

Assume $\mathcal{S}$ is the largest set containing a given $m_1'$, and that $d\gcd(\tilde n_2, n_1)$ is the smallest non-zero difference between two elements in $\mathcal{S}$. There is then an $m_1 \in \mathcal{S}$ such that $\gcd(m_1 \tilde n_2, n_1) = d \gcd(\tilde n_2, n_1)$. By dividing with $\gcd(\tilde n_2, n_1)$ and using the relation $\gcd(b,c)=1\Rightarrow\gcd(ab,c)=\gcd(a,c)$ (Sec.\ \ref{sect:b_begone}), an equivalent expression is $d=\gcd(m_1, n_1/\gcd(\tilde n_2, n_1))$. An integer $t$ then exists, such that
\begin{equation}
	m_1 = m_1' + t d \gcd(\tilde n_2, n_1) \mod n_1.
\end{equation}
Then,
\begin{equation}
	d \gcd(\tilde n_2, n_1) = \gcd(m_1 \tilde n_2, n_1) = \gcd\left((m_1' + t d \gcd(\tilde n_2, n_1)) \tilde n_2, n_1\right).
\end{equation}
Dividing both sides by $\gcd(\tilde n_2, n_1)$ yields
\begin{equation}
	d = \gcd\left(m_1' + t d \gcd(\tilde n_2, n_1), \frac{n_1}{\gcd(\tilde n_2, n_1)}\right).
\end{equation}
It is shown in Sec.\ \ref{sect:t_exists} that a value of $t$ exists -- the value corresponding to the desired $m_1$ -- such that
\begin{equation}\label{eq:in_set_condition}
	d = \gcd\left(m_1', d \gcd(\tilde n_2, n_1), \frac{n_1}{\gcd(\tilde n_2, n_1)}\right).
\end{equation}
When this relation is fulfilled, $m_1'$ is part of such a set $\mathcal{S}$.

It is then shown that $m_1'$ is included in a set with $d$ given as in Eq.\ (25) in the main text, which satisfies Eq.\ \eqref{eq:in_set_condition}. By Eq.\ (25),
\begin{equation}
	\gcd\left(m_1', \frac{n_1}{\gcd(\tilde n_2, n_1)} \right) = \gcd\left(m_1', v\right)\gcd\left(m_1', r\right) = d \gcd(m_1', r) \quad \text{with }\gcd(r, \tilde n_2) = 1.
\end{equation}
Inserting in Eq.\ \eqref{eq:in_set_condition},
\begin{equation}
	\begin{split}
		\gcd\left(m_1', d \gcd(\tilde n_2, n_1), \frac{n_1}{\gcd(\tilde n_2, n_1)}\right) &= 
		\gcd\left(d \gcd(\tilde n_2, n_1), d\gcd(m_1', r)\right) \\
			&= d\gcd\left( \tilde n_2, n_1, m_1', r\right) \\
			&= d\gcd\left( \gcd(\tilde n_2, r), n_1, m_1'\right)
			= d.
	\end{split}
\end{equation}

It finally needs to be shown that Eq.\ \eqref{eq:in_set_condition} does not hold for any other $d'$ where the corresponding $\mathcal{S}'$ is larger than the $\mathcal{S}$ corresponding to $d$, i.e.\ $\mathcal{S}$ is the largest set containing $m_1'$. Sets $\mathcal{S}'$ smaller than $\mathcal{S}$ are irrelevant, since the largest possible $\mathcal{S}$ is desired. The set $\mathcal{S}'$ can only be larger than $\mathcal{S}$ if $d>d'$. Since $d'=\gcd(m_1, n_1/\gcd(\tilde n_2, n_1))$ by assumption, $d'$ must be a divisor of $m_1$ and $n_1/\gcd(\tilde n_2, n_1)$. It is always possible to partition $d'$ as $d'=ws$ where $w\mid v$ and $s\mid r$, with $v$ given by Eq.\ (25) in the main text and $vr=n_1/\gcd(\tilde n_2, n_1)$. As $w$ also divides $m_1'$, $w\mid d$. Inserting in Eq.\ \eqref{eq:in_set_condition} and dividing by $d'$,
\begin{equation}
	\begin{split}
		\gcd\left( \frac{m_1'}{d'}, \gcd(\tilde n_2, n_1), \frac{n_1}{d' \gcd(\tilde n_2, n_1)} \right) 
		&= \gcd\left( \frac{m_1'}{d'}, \tilde n_2, n_1, \frac{n_1}{d' \gcd(\tilde n_2, n_1)} \right) \\
			&= \gcd\left( \frac{m_1'}{d'}, \tilde n_2, \frac{n_1}{d' \gcd(\tilde n_2, n_1)} \right) \\
			&= \gcd\left( \frac{m_1'}{ws}, \tilde n_2, \frac{vr}{ws} \right) \\
			&= \gcd\left( \frac{m_1'}{w}, \tilde n_2, \frac{v}{w} \right) \\
			&= \gcd\left( \frac{d}{w}, \tilde n_2 \right).
	\end{split}
\end{equation}
With $d$ being a divisor of $v$, $d/w$ must be as well. Then $d/w$ is certainly not coprime to $v$. This further implies that it is not coprime to $\tilde n_2$, i.e.\
\begin{equation}
		\gcd\left( m_1', d'\gcd(\tilde n_2, n_1), \frac{n_1}{\gcd(\tilde n_2, n_1)} \right) 
			= d'\gcd\left( \frac{d}{w}, \tilde n_2 \right) > d',
\end{equation}
and the relation in Eq.\ \eqref{eq:in_set_condition} is not fulfilled, as was to be shown.

\section{Proofs for the Three-Dimensional Algorithm}

\subsection{Step 1}\label{sect:proof_3d_step_1}

Start at the point $(m_1, m_2, m_3)$. The size of the group $\langle m_3 \rangle_{n_3}$ is $n_3/\gcd(m_3, n_3)$, so it takes that many multiples of the starting point to loop around the circle of tori. Looping $t$ times and adding $k$ more multiples of $(m_1, m_2, m_3)$, the starting point becomes $(m_1', m_2', m_3')$, with
\begin{equation}
	\begin{alignedat}{2}
		m_3' &= m_3\left( k + t \frac{n_3}{\gcd(m_3, n_3)} \right) = m_3 k &&\mod n_3, \\
		m_2' &= m_2\left( k + t \frac{n_3}{\gcd(m_3, n_3)} \right) &&\mod n_2, \\
		m_1' &= m_1\left( k + t \frac{n_3}{\gcd(m_3, n_3)} \right) &&\mod n_1.
	\end{alignedat}
\end{equation}
The possible values of $m_i'$, $i=1,2$, are
\begin{equation}
	m_i' \in \left\langle k + t \frac{n_3}{\gcd(m_3, n_3)} \right\rangle_{n_i} = \left\langle \gcd\left(k + t \frac{n_3}{\gcd(m_3, n_3)}, n_i\right) \right\rangle_{n_i}.
\end{equation}
Torus $m_3'$ is then covered if and only if there exists a value of $t$ such that
\begin{equation}
	\gcd\left(k + t \frac{n_3}{\gcd(m_3, n_3)}, n_i\right) = 1
\end{equation}
for both $i=1$ and $2$, which is the case if and only if
\begin{equation}
	\gcd\left(k + t \frac{n_3}{\gcd(m_3, n_3)}, n_1 n_2\right) = 1,
\end{equation}
see Sec.\ \ref{sect:gcd_set_of_ints}. By the property in Sec.\ \ref{sect:t_exists}, a value of $t$ satisfying the above exists if and only if
\begin{equation}
	\gcd\left(k, \frac{n_3}{\gcd(m_3, n_3)}, n_1 n_2\right) = 1.
\end{equation}

Since $m_3$ is a divisor of $v=\pi_{n_1n_2}(n_3)$, it is also a divisor of $n_3$, i.e.\ $\gcd(m_3, n_3)=m_3$. By the assumption $m_3 = \gcd(m_3', v)$, $m_3$ divides $m_3'$ so $k=m_3'/m_3$. The proof then follows directly from Sec.\ \ref{sect:proof_2d_step_1} by replacing in that section $n_2$ with $n_3$, $m_2$ with $m_3$, and $n_1$ with $n_1n_2$.

\subsection{Step 2}\label{sect:proof_3d_step_2}

The proof follows directly from the results in Sec.\ \ref{sect:proof_2d_step_2}, by replacing in that section $n_1$ with $n_2$, $\tilde n_2$ with $\tilde n_3$, and $m_1$ with $m_2$.

\subsection{Step 3}\label{sect:proof_3d_step_3}

For every $m_2''\in\mathcal{S}$, it follows from Eq.\ (29) in the main text that there exists an integer $f$, such that
\begin{equation}\label{eq:m2pp}
	\gcd\left( m_2'', \frac{n_2}{\gcd(\tilde n_3, n_2)} \right) = df = der, \quad \text{where }~e = \pi_{n_1}(f)~\text{ and } \gcd(n_1, r) = 1.
\end{equation}
Since $f$ is a divisor of $\tilde n_2/\gcd(\tilde n_3, n_2)$,
\begin{equation}
	e = \pi_{n_1}(f) ~~\Big\vert~~ \pi_{n_1}\left( \frac{\tilde n_2}{\gcd(\tilde n_3, n_2)} \right) = w,
\end{equation}
which follows from the property in Sec.\ \ref{sect:pia_pib}. It must be shown that every $m_2''$ of this form is covered by a single $m_2'\in\mathcal{S}$, which has
\begin{equation}\label{eq:m2p}
	\gcd\left( m_2', \frac{n_2}{\gcd(\tilde n_3, n_2)} \right) = de \quad \Longleftrightarrow \quad \gcd\left( \frac{m_2'}{de}, \frac{\dtilde n_2}{\gcd(\tilde n_3, n_2)} \right) = 1
\end{equation}
where $\dtilde n_2 = n_2/ed$.

It is first shown that a non-empty subset $\mathcal{T}\subseteq\mathcal{S}$ exists for every divisor $e$ of $w$. Assume $m_2\in\mathcal{S}$ reaches every $m_2'\in\mathcal{S}$, so
\begin{equation}
	m_2' = m_2(1+x\tilde n_3)\mod n_2 \quad \text{and} \quad \gcd\left( m_2, \frac{n_2}{\gcd(\tilde n_3, n_2)} \right) = d.
\end{equation}
Then,
\begin{equation}
	\begin{split}
		\gcd\left( m_2', \frac{n_2}{\gcd(\tilde n_3, n_2)} \right) &= \gcd\left( m_2(1+x\tilde n_3), \frac{n_2}{\gcd(\tilde n_3, n_2)} \right) \\
			&= d\gcd\left( \frac{m_2}{d}(1+x\tilde n_3), \frac{\tilde n_2}{\gcd(\tilde n_3, n_2)} \right) \\
			&= d\gcd\left( 1+x\tilde n_3, \frac{\tilde n_2}{\gcd(\tilde n_3, n_2)} \right) \\
			&= d\gcd\left( 1+x\tilde n_3, \tilde n_2 \right).
	\end{split}
\end{equation}
For all $x$, the set of values taken by $\gcd\left( 1+x\tilde n_3, \tilde n_2 \right)$ are (see Sec.\ \ref{sect:possible_values})
\begin{equation}
	\{ \gcd(1+x\tilde n_3, \tilde n_2) : x \in \mathbb{Z} \} = \text{ divisors of }~\frac{\tilde n_2}{\pi_{\tilde n_3}(\tilde n_2)}.
\end{equation}
Every divisor $e$ of $w$ is also a divisor of $\tilde n_2 / \pi_{\tilde n_3}(\tilde n_2)$, so for every divisor $e$ of $w$ an $x$ exists such that $\gcd(1+x\tilde n_3, \tilde n_2) = e$. Therefore, for every divisor $e$ of $w$, an $m_2'\in\mathcal{S}$ exists satisfying Eq.\ \eqref{eq:m2p}. As $ed$ divides this $m_2'$, $m_2'\in \mathcal{T}$, i.e.\ the $\mathcal{T}$ corresponding to $e$ is non-empty.

It is then shown that every $m_2''\in\mathcal{T}\subseteq\mathcal{S}$ can be reached by an $m_2'\in\mathcal{T}\subseteq\mathcal{S}$ satisfying Eq.\ \eqref{eq:m2p}. This is done by explicitly showing that an $x'$ exists, such that $m_2'' = m_2'(1+x'\tilde n_3)\smod n_2$. Assume $m_2\in\mathcal{S}$ reaches every element of $\mathcal{S}$. Let $c$ be the smallest integer in $\mathcal{S}$. Since the smallest non-zero difference between two elements in $\mathcal{S}$ is $d\gcd(\tilde n_3, n_2)$, there exist integers $a$ and $b$ such that
\begin{equation}
	\begin{split}
		m_2' &= c + ad\gcd(\tilde n_3, n_2), \\
		m_2'' &= c + bd\gcd(\tilde n_3, n_2), \\
			&\Longrightarrow\quad m_2'' - m_2' = (b-a)d\gcd(\tilde n_3, n_2).
	\end{split}
\end{equation}
The left-hand side is divisible by $e$ since $m_2',m_2''\in\mathcal{T}$, and by extension the right-hand side is divisible by $e$ as well. However, since $e$ is a divisor of $w$ (as given in Eq.\ (31) in the main text), and since $\tilde n_2 / \pi_{\tilde n_3}(\tilde n_2)$ must be coprime to $\tilde n_3$ by definition of the $\pi$ function, $e$ must be coprime to $\tilde n_3$ as well. Property \ref{sect:pi_coprime_coprime} then implies that $e$ is coprime to $v=\pi_{\tilde n_3}( n_2/\gcd(\tilde n_3, n_2))$, and thereby to $d$, since $d\mid v$ by assumption. Then $\gcd(e, d\gcd(\tilde n_3, n_2)) = 1$, so (by the property in Sec.\ \ref{sect:c_divides_ab})
\begin{equation}
	\frac{m_2''-m_2'}{e} = \frac{(b-a)d\gcd(\tilde n_3, n_2)}{e} = \frac{b-a}{e}d\gcd(\tilde n_3, n_2) \quad \Longrightarrow \quad ed\gcd(\tilde n_3, n_2) \mid m_2'' - m_2'.
\end{equation}
Furthermore, there exists an integer $k$, such that
\begin{equation}
	\begin{split}
		\gcd&(m_2'\tilde n_3, n_2) = k\gcd(\tilde n_3, n_2) \\
		&\Longrightarrow \quad k = \gcd\left(m_2'\frac{\tilde n_3}{\gcd(\tilde n_3, n_2)}, \frac{n_2}{\gcd(\tilde n_3, n_2)}\right) = \gcd\left(m_2', \frac{n_2}{\gcd(\tilde n_3, n_2)}\right) = de.
	\end{split}
\end{equation}
The last equality follows from the assumption Eq.\ \eqref{eq:m2p}. Consequently, 
\begin{equation}
	\langle m_2' \tilde n_3 \rangle_{n_2} = \langle \gcd(m_2' \tilde n_3, n_2) \rangle_{n_2} =  \langle ed\gcd(\tilde n_3, n_2) \rangle_{n_2},
\end{equation}
implying that for any integer $t$, there exists an integer $x'$ such that $x'm_2'\tilde n_3 = ted\gcd(\tilde n_3, n_2)\smod n_2$. Choosing $t=(m_2''-m_2')/ed\gcd(\tilde n_3, n_2)$, it is found that
\begin{equation}
	m_2' (1 + x'\tilde n_3) = m_2' + x' m_2' \tilde n_3 = m_2' + t ed\gcd(\tilde n_3, n_2) = m_2' + m_2''-m_2' = m_2'' \mod n_2
\end{equation}
which was to be shown.

Next it is shown that $m_2'$ covers $m_2''$. Since it reaches $m_2''$, this can be written as
\begin{equation}
	m_2'' = m_2'(1+x'\tilde n_3) \mod n_2.
\end{equation}
Inserting in Eq.\ \eqref{eq:m2pp},
\begin{equation}
	\begin{split}
		\gcd\left( m_2'', \frac{n_2}{\gcd(\tilde n_3, n_2)} \right) &= de\gcd\left( \frac{m_2''}{de}, \frac{\dtilde n_2}{\gcd(\tilde n_3, n_2)} \right) \\
			&= de\gcd\left( \frac{m_2'}{de}(1+x'\tilde n_3), \frac{\dtilde n_2}{\gcd(\tilde n_3, n_2)} \right) \\
			&= de\gcd\left( 1+x'\tilde n_3, \frac{\dtilde n_2}{\gcd(\tilde n_3, n_2)} \right) \\
			&= de\gcd\left( 1+x'\tilde n_3, \dtilde n_2 \right) \\
			&\Longrightarrow\quad \gcd(1+x'\tilde n_3, \dtilde n_2) = r.
	\end{split}
\end{equation}

Starting from $m_2'$ it takes $\lcm(\tilde n_3, \tilde n_2) / e = \lcm(\tilde n_3, \dtilde n_2)$ (see Sec.\ \ref{sect:lcm_div}) multiples of $m_2'$ to loop around the circle of tori. Looping $t$ times yields the first coordinate
\begin{equation}\label{eq:m1pp}
	\begin{split}
		m_1'' = m_1'(1+x'\tilde n_3 + t\lcm(\tilde n_3, \dtilde n_2)) \mod n_1.
	\end{split}
\end{equation}

Note that
\begin{equation}
	\gcd(1+x'\tilde n_3,\lcm(\tilde n_3, \dtilde n_2)) = \gcd\left(1+x'\tilde n_3,\frac{\tilde n_3 \dtilde n_2}{\gcd(\tilde n_3, \dtilde n_2)}\right) = \gcd\left(1+x'\tilde n_3,\dtilde n_2\right).
\end{equation}
For different starting points $m_1'$ on circle $m_2'$, the different values $m_1''$ can take are
\begin{equation}\label{eq:m1pp_can_take}
	m_1'' \in \left\langle 1+x'\tilde n_3 + t\lcm(\tilde n_3, \dtilde n_2) \right \rangle_{n_1} = \left\langle \gcd(1+x'\tilde n_3, \dtilde n_2, n_1) \right \rangle_{n_1} = \left\langle \gcd(r, n_1) \right \rangle_{n_1} = \langle 1 \rangle_{n_1} = \mathbb{Z}_{n_1}.
\end{equation}
This follows from the property in Sec.\ \ref{sect:t_exists}, and implies that $m_2''$ is covered by $m_2'$.

Let $m_2'''\in \mathcal{T}$ with
\begin{equation}
	\gcd\left(m_2''', \frac{n_2}{\gcd(\tilde n_3, n_2)} \right) = des, \quad \gcd(s, n_1) = 1.
\end{equation}
It is shown that every $m_2'''$ covers the same $m_2''$ as $m_2'$. If $m_2'''$ does not reach $m_2''$, it certainly does not cover it, so assume $m_2'''$ reaches $m_2''$, i.e.\ $m_2'' = m_2'''(1+x'\tilde n_3)\smod n_2$. Then,
\begin{equation}
	\begin{split}
		\gcd\left( m_2'', \frac{n_2}{\gcd(\tilde n_3, n_2)} \right) &= des\gcd\left( \frac{m_2''}{des}, \frac{\dtilde n_2}{s\gcd(\tilde n_3, n_2)} \right) \\
			&= des\gcd\left( \frac{m_2'''}{des}(1+x'\tilde n_3), \frac{\dtilde n_2}{s\gcd(\tilde n_3, n_2)} \right) \\
			&= des\gcd\left( 1+x'\tilde n_3, \frac{\dtilde n_2}{s\gcd(\tilde n_3, n_2)} \right) \\
			&= de\gcd\left( s(1+x'\tilde n_3), \dtilde n_2 \right) \\
			&\Longrightarrow\quad \gcd\left(s(1+x'\tilde n_3), \dtilde n_2\right) = r.
	\end{split}
\end{equation}
Inserting in the gcd in Eq.\ \eqref{eq:m1pp_can_take},
\begin{equation}
	\gcd(1+x'\tilde n_3, \dtilde n_2, n_1) = \gcd(s(1+x'\tilde n_3), \dtilde n_2, n_1) = \gcd(r, n_1).
\end{equation}
This is the same result as in Eq.\ \eqref{eq:m1pp_can_take}, so any $m_2'''$ in this form covers $m_2''$ if and only if $m_2'$ covers it.

It is finally shown that $m_2''$ on the form given in Eq.\ \eqref{eq:m2pp} is not covered by $m_2'$ with
\begin{equation}
	\gcd\left( m_2', \frac{n_2}{\gcd(\tilde n_3, n_2)} \right) = de' < de.
\end{equation}
If $m_2'$ does not reach $m_2''$, it certainly does not cover it, so assume $m_2'$ reaches $m_2''$, i.e.\ $m_2'' = m_2'(1+x'\tilde n_3)\smod n_2$. This implies that $e'\mid m_2''$ and thereby $e'\mid e$, as may be seen by dividing both sides of Eq.\ \eqref{eq:m2pp} by $de'$, and noting that since $e'\mid w$ it must be coprime to $r$. The corresponding definition of $\dtilde n_2$ is in this case $\dtilde n_2 = n_2 / de'$, so
\begin{equation}
	\begin{split}
		\gcd\left( m_2'', \frac{n_2}{\gcd(\tilde n_3, n_2)} \right) &= de'\gcd\left( \frac{m_2''}{de'}, \frac{\dtilde n_2}{\gcd(\tilde n_3, n_2)} \right) \\
			&= de'\gcd\left( \frac{m_2'}{de'}(1+x'\tilde n_3), \frac{\dtilde n_2}{\gcd(\tilde n_3, n_2)} \right) \\
			&= de'\gcd\left( 1+x'\tilde n_3, \frac{\dtilde n_2}{\gcd(\tilde n_3, n_2)} \right) \\
			&= de'\gcd\left( 1+x'\tilde n_3, \dtilde n_2 \right) \\
			&\Longrightarrow\quad \frac{e}{e'}r = \gcd\left( 1+x'\tilde n_3, \dtilde n_2 \right).
	\end{split}
\end{equation}
Inserting this in the gcd in Eq. \eqref{eq:m1pp_can_take},
\begin{equation}
	\gcd(1+x'\tilde n_3, \dtilde n_2, n_1) = \gcd\left(\frac{e}{e'}, n_1\right) > 1.
\end{equation}
The final inequality follows since $e\mid w$ and $e'\mid w$, which implies $e/e' \mid w$, so $e/e'$ cannot be coprime to $n_1$. This circle $m_2'$ does in conclusion not cover $m_2''$.

\section{The Number of \textbf{q}-points on a Prime Grid}\label{sect:prime_grids}

Assume $n$ is a prime number. It must be proven that the number of supercells required for an $n\times n$ or $n\times n \times n$ $\mathbf{q}$-grid, i.e.\ the size of $\mathcal{Q}$, is
\begin{equation}
	\begin{alignedat}{2}
		N_\text{sc} &= n + 1 \quad &&(\text{two dimensions}), \\
		N_\text{sc} &= n^2 + n + 1 \quad &&(\text{three dimensions}).
	\end{alignedat}
\end{equation}

\subsection{Two Dimensions}
Starting at the point $(m_1, 1)$, the point $(m_1m_2, m_2)$ is covered. As all the different starting points on circle $1$ are considered, the points covered on circle $m_2$ have the first coordinates
\begin{equation}
	m_1m_2 \in \langle m_2 \rangle_n = \langle \gcd(m_2, n) \rangle_n = 
	\begin{cases}
		\mathbb{Z}_n	&\text{for }m_2 \neq 0\smod{n} \\
		\{0\} 			&\text{for }m_2 = 0\smod{n}
	\end{cases}
\end{equation}
where it has been used that $n$ is prime, so $\gcd(m_2,n) = 1$ or $n$. Circle 1 thereby covers every circle except $m_2 = 0$, where every starting point becomes $(0, 0)$ and no other point on this circle. For $m_2=n+1$ the starting point becomes $(m_1(n+1), n+1) = (m_1, 1)\smod(n, n)$, so every point on circle 1 covers exactly one point on each circle. This implies that every $\mathbf{q}=(m_1, 1)$ for $m_1=0,\dots,n-1$ must be added to $\mathcal{Q}$, bringing the subtotal size of $\mathcal{Q}$ to $n$. The points which have so far not been covered are all the points $(m_1, 0)$ for $m_1 \neq 0$. These are all covered by $\mathbf{q}=(1, 0)$ which must be added to $\mathcal{Q}$ as well. All points on the torus are now covered, so the total size of $\mathcal{Q}$ is $n+1$.
{\hfill$\blacksquare$}

\subsection{Three Dimensions}

The proof in three dimensions proceeds in much the same way as in two dimensions. Start at the point $(m_1, m_2, 1)$ which covers $(m_1m_3, m_2m_3, m_3)$. On torus $m_3$, the first and second coordinates can take the values
\begin{equation}
	m_im_3 \in \langle m_3 \rangle_n = \langle \gcd(m_3, n) \rangle_n = 
	\begin{cases}
		\mathbb{Z}_n	&\text{for }m_3 \neq 0\smod{n} \\
		\{0\} 			&\text{for }m_3 = 0\smod{n}
	\end{cases}
\end{equation}
Torus 1 then covers all other tori except $m_3 = 0$, where only $(0, 0, 0)$ is covered. At $m_3=n+1$ the starting point becomes $(m_1(n+1), m_2(n+1), n+1) = (m_1, m_2, 1)\smod (n, n, n)$ implying that each starting point on torus 1 covers exactly one point on one circle on each torus. Every possible starting point $\mathbf{q}=(m_1, m_2, 1)$ on torus 1 must then be added to $\mathcal{Q}$, which consequently has a subtotal size of $n^2$. The points not covered so far are all the points on torus 0 except $(0, 0, 0)$. It was proven above that covering a single torus takes $n+1$ starting points, which must all be added to $\mathcal{Q}$ as well. The final size of $\mathcal{Q}$ is thereby $n^2+n+1$.
{\hfill$\blacksquare$}

\clearpage

\section{Complete Timings for All Calculations}\label{sect:timings}
\setlength{\tabcolsep}{10pt}
\centering
\begin{minipage}[t]{0.45\linewidth}
	\centering
	\textbf{Al} \\
	\vspace*{5pt}
	\begin{tabular}{| c | c | c | c |}
		\hline
		$n$ & Cover & Non-diag. & Diag. \\
		\hline
		2   & 0.59 & 0.70 & 0.21 \\
		4   & 4.0 & 4.9 & 2.0 \\
		6   & 17.2 & 20.1 & 18.7 \\
		7   & 13.1 & & \\
		8   & 28.3 & 60.6 & 286.8 \\
		10  & 64.5 & 132.6 & 707.8 \\
		11  & 44.9 & & \\
		12  & 121.6 & 260.5 & \\
		13  & 74.3 & & \\
		14  & 154.1 & 503.5 & \\
		16  & 210.6 & 906.1 & \\
		17  & 161.4 & & \\
		\hline
	\end{tabular} \\
	\vspace*{5pt}
	CPU hours for an $n\times n\times n$ grid.

	\normalsize
	\vspace*{15pt}
	\textbf{PtS} \\
	\vspace*{5pt}
	\begin{tabular}{| c | c | c | c |}
		\hline
		$n$ & Cover & Non-diag. & Diag. \\
		\hline
		1  & 2.3 & 2.7 & 1.5 \\
		2  & 26.8 & 29.9 & 193.5 \\
		3  & 177.3 & 207.8 & 1638 \\
		4  & 507.3 & 1023 & 5458 \\
		5  & 1141 & 3485 & 34573* \\
		6  & 4087 & 9051 &  \\
		7  & 4494 & 19538 &  \\
		8  & 9950 & 44847 &  \\
		\hline
	\end{tabular} \\
	\vspace*{5pt}
	CPU hours for a $2n\times 2n\times n$ grid.
	
	*Estimate.
\end{minipage}
\begin{minipage}[t]{0.45\linewidth}
	\centering
	\textbf{CaF\textsubscript{2}} \\
	\vspace*{5pt}
	\begin{tabular}{| c | c | c | c |}
		\hline
		$n$ & Cover & Non-diag. & Diag. \\
		\hline
		2  & 2.0 & 2.2 & 1.0 \\
		4  & 13.8 & 15.3 & 91.2 \\
		6  & 71.3 & 72.6 & 1034 \\
		8  & 131.3 & 267.3 & 3304 \\
		10  & 349.9 & 770.5 & \\
		12  & 841.3 & 2135 & \\
		14  & 1237 & 4899 & \\
		16  & 1839 & 10240 & \\
		\hline
	\end{tabular} \\
	\vspace*{5pt}
	CPU hours for an $n\times n\times n$ grid.

	\normalsize
	\vspace*{15pt}
	\textbf{Nb\textsubscript{3}Sn} \\
	\vspace*{5pt}
	\begin{tabular}{| c | c | c | c |}
		\hline
		$n$ & Cover & Non-diag. & Diag. \\
		\hline
		2  & 81.9 & 87.0 & 143.2 \\
		4  & 1080 & 1125 & 11298 \\
		\hline
	\end{tabular} \\
	\vspace*{5pt}
	CPU hours for an $n\times n\times n$ grid.

	\normalsize
	\vspace*{15pt}
	\textbf{Graphene} \\
	\vspace*{5pt}
	\begin{tabular}{| c | c | c | c |}
		\hline
		$n$ & Cover & Non-diag. & Diag. \\
		\hline
		3  & 0.22 & 0.23 & 0.15 \\
		6  & 1.2 & 1.5 & 1.0 \\
		9  & 1.8 & 5.9 & 6.8 \\
		12  & 4.9 & 12.5 & 12.2 \\
		15  & 6.8 & 27.0 & 44.5 \\
		18  & 12.8 & 47.0 & 124.6 \\
		\hline
	\end{tabular} \\
	\vspace*{5pt}
	CPU hours for an $n\times n$ grid.
\end{minipage}

\vspace*{1cm}

\bibliographystyle{unsrt}
\bibliography{si}